\documentclass[fleqn,usenatbib]{mnras}
\usepackage{fix-cm}

\usepackage{newtxtext,newtxmath}
\usepackage{float}
\usepackage{placeins}
\usepackage{capt-of}
 \usepackage{lmodern}

\usepackage[T1]{fontenc}
\usepackage{xcolor}
\usepackage{needspace}
\usepackage{hyperref}

\DeclareRobustCommand{\VAN}[3]{#2}
\let\VANthebibliography\thebibliography
\def\thebibliography{\DeclareRobustCommand{\VAN}[3]{##3}\VANthebibliography}

\usepackage{graphicx}	
\usepackage{amsmath}	

\title[AGE-PRO: Multi-wavelength Analysis]{The ALMA Survey of Gas Evolution of PROtoplanetary disks (AGE-PRO): Evolution of dust properties from multi-wavelength observations}

\author[]{
Camila Pulgar\'es$^{1,2}$\thanks{E-mail: camila.pulgares.25@ucl.ac.uk},
Laura M. P\'erez$^{1}$,
Carolina Agurto-Gangas$^{3}$,
Lilian Luo$^{2}$,
Paola Pinilla$^{2}$,
\newauthor
Anibal Sierra$^{2,4}$,
Miguel Vioque$^{5}$,
Nicol\'as Kurtovic$^{6,7}$,
Rossella Anania$^{8}$, 
John Carpenter$^{9}$, 
Lucas Cieza$^{10,11}$,
\newauthor
Dingshan Deng$^{12}$, 
Camilo González-Ruilova$^{13, 14, 15}$,
Michiel R. Hogerheijde$^{16,17}$,
James Miley$^{9,13,14,15,18}$,
\newauthor
Ke Zhang$^{19}$
\\
$^{1}$Departamento de Astronom\'ia, Universidad de Chile, Camino El Observatorio 1515, Las Condes, Santiago, Chile\\
$^{2}$Mullard Space Science Laboratory, University College London, Holmbury St Mary, Dorking, Surrey RH5 6NT, UK\\
$^{3}$Departamento de F\'isica, Universidad T\'ecnica Federico Santa Mar\'ia, Vicuña Mackena 3939, San Joaqu\'in, Santiago de Chile, Chile\\
$^{4}$Universidad Nacional Autónoma de México, Instituto de Astronomía, A.P. 70-264, Ciudad de México 04510, México\\
$^{5}$European Southern Observatory, Karl-Schwarzchild-Str. 2, 85748 Garching bei München, Germany\\
$^{6}$Max Planck Institute for Extraterrestial Physics, Giessenbachstrasse 1, D-85748 Garching, Germany\\
$^{7}$Max-Planck-Institut fur Astronomie (MPIA), Konigstuhl 17, 69117 Heidelberg, Germany\\
$^{8}$School of Physics, Trinity College Dublin, the University of Dublin, College Green, Dublin 2, Ireland\\
$^{9}$Joint ALMA Observatory, Alonso de C\'ordova 3107, Vitacura, Santiago 763-0355, Chile\\
$^{10}$Instituto de Estudios Astrofísicos, Universidad Diego Portales, Av. Ejercito 441, Santiago, Chile\\
$^{11}$Millennium Nucleus on Young Exoplanets and their Moons (YEMS), Chile\\
$^{12}$Lunar and Planetary Laboratory, the University of Arizona, Tucson, AZ 85721, USA\\
$^{13}$Departamento de Física, Universidad de Santiago de Chile, Avenida Victor Jara 3659, Santiago, Chile\\
$^{14}$Millenium Nucleus of Young Exoplanets and their Moons (YEMS), Chile\\
$^{15}$Center for Interdisciplinary Research in Astrophysics and Space Exploration (CIRAS), Universidad de Santiago, Chile\\
$^{16}$Leiden Observatory, Leiden University, Einsteinweg 55, 2333 CC Leiden, The Netherlands\\
$^{17}$Anton Pannekoek Institute for Astronomy, University of Amsterdam, The Netherlands\\
$^{18}$European Southern Observatory, Alonso de Córdova 3107, Vitacura, Santiago, Chile\\
$^{19}$Department of Astronomy, University of Wisconsin-Madison, 475 N Charter St, Madison, WI 53706, USA\\}

\date{Accepted 2026 October 22. Received 2026 October 22; in original form 2026 May 21}

\pubyear{\the\year{}}

\begin{document}
\label{firstpage}
\pagerange{\pageref{firstpage}--\pageref{lastpage}}
\maketitle

\begin{abstract}
Understanding the evolution of solids in protoplanetary discs is important to comprehend the formation of giant planets cores and terrestrial planets. The ALMA Large Program AGE-PRO studied key disc properties in a survey of 30 protoplanetary discs across three evolutionary stages: the embedded phase (0.5-1 Myr), the middle-age phase (1-3 Myr), and the end of the disc lifetime (2-6 Myr). Preliminary dust properties were estimated using 1.3 mm observations, but a multi-wavelength analysis provides a better constraint of important properties such as the dust disc mass, the dust temperature, and the maximum grain size. We present new ALMA 3.1 mm and 1.5 mm (Bands 3 and 5) continuum observations of the complete survey of 30 AGE-PRO discs. We fitted the observed visibilities and derived radial brightness profiles, spatially resolving 20 discs from the AGE-PRO sample at the two studied wavelengths. A multi-wavelength modelling was performed including also the radial profiles from the 1.3 mm observations. Our results replicate trends obtained from the single wavelength analysis: the dust disc size seems to be constant over age and the dust disc mass decreases with time, with a factor $\sim$10-40 uncertainty from opacity assumptions. This result implies much lower gas-to-dust mass ratios for the discs in our sample than previously reported. In addition, the derived spectral indices are consistent with optically thick emission in a large fraction of the AGE-PRO discs, in particular for the Ophiuchus and Lupus discs. 
\end{abstract}




\section{Introduction}

Planet formation takes place in discs of dust and gas that orbit a young protostar, as a consequence of the conservation of angular momentum after the collapse of a molecular cloud. Although these discs are mainly composed of gas, with only a few percent of the total mass coming from  dust, the solid component is crucial in the formation of planetary bodies: dust grains serve as the building blocks for pebbles and planetesimals, influencing the efficiency of core accretion and the composition of forming planets \citep{testi2014, william+cieza2011, drazkowska2023}. 

A detailed understanding of these physical processes not only allows us to reconstruct the early stages of planet formation, but also provides essential clues to explain the large diversity observed among exoplanet systems. More than 6000 exoplanets have been discovered in past years, showing that planetary systems differ widely from what we observe in the Solar system \citep{2015ARA&A..53..409W}. Variations in the structure and evolution of protoplanetary discs can lead to different outcomes of the properties of planets \citep[e.g.,][]{andrews2020}. Therefore, studying the physics of discs and the processes governing the assembly of pebbles, planetesimals, and planetary cores is crucial for linking the initial conditions and evolution in planet-forming discs.

High-resolution imaging from the Atacama Large Millimeter/submillimeter Array (ALMA) and near-infrared instruments like VLT/SPHERE have provided unprecedented insights into the complex morphologies of these discs \citep{alma2015, nienke2017, garufi2020, benisty2023}. Observations have revealed that substructures such as rings, gaps, central cavities, spirals, and azimuthal asymmetries/arcs are ubiquitous in protoplanetary discs \citep[e.g.,][]{alma2015, perez2016, andrews2018, andrews2020, bae2023}. These features have been detected in multiple systems and are often interpreted as signatures of unseen young planets, in the early stages of their formation \citep[e.g.,][]{zhu2011,pinilla2012,rosotti2016}, or as  signatures of magneto‐hydrodynamic or gravitational instabilities \citep[e.g.,][]{forgan2018}, which can serve as dust-trapping  mechanisms that regulate the growth and radial drift of dust grains \citep{pinilla2012,testi2014}. 

In a protoplanetary disc lacking dust traps (i.e., a smooth disc), particles with sizes in the mm–cm range are rapidly lost due to radial drift, on timescales significantly shorter than the disc lifetime \citep[<0.1 Myr,][]{brauer2008}. An effective mechanism to mitigate this process is dust trapping at local gas pressure maxima \citep[e.g.,][]{pinilla2012}, as particles tend to migrate toward regions of higher pressure. Solids can be efficiently retained if the gas pressure profile exhibits localized enhancements. Within these pressure bumps, radial drift is halted, collision velocities decrease, and particle growth can proceed further.

Different dust properties, such as the dust disc mass, have been derived for a large number of discs in several star-forming regions \citep[e.g.,][]{andrews2013, ansdell2016, pascucci2016, barenfeld2016, 2019ApJ...875L...9W, 2025MNRAS.543.2723P} using single-wavelength observations and adopting a disc-averaged temperature and dust opacity and also assuming that the observed emission is optically thin \citep[see, e.g.,][]{Tripathi2017, tazzari2021}. For moderately optically thick discs, these assumptions can lead to large underestimation of the total solid mass in discs \citep{liu2022}. \cite{viscardi2025} showed that a (sub)mm SED analysis is significantly more robust than assuming optically thin emission at a single wavelength and that spatial resolution does not critically affect the derived dust mass estimates. However, it is necessary to include optically thinner observations, typically done by adding longer wavelength data in the analysis. Multi-wavelength analysis has also allowed to constrain the maximum grain size in discs. Together with the dust surface density, these properties can be used to test the presence of dust traps in these systems, as a localized peak is expected at the trap location for these two properties, as was done in \cite{sierra2021}.

More recently, the ALMA Large Program AGE-PRO (“ALMA survey of Gas Evolution in PROtoplanetary discs”) observed 30 protoplanetary discs of different ages, ranging from 0.1 to 6 Myr, to explore how the gas emission evolves through the lifetime of the discs \citep{2025ApJ...989....1Z}. The sample is composed by 10 young class I/ Flat Spectrum discs from the Ophiuchus region \citep{2025ApJ...989....2R}, 10 middle-aged class II discs from the Lupus region \citep{2025ApJ...989....3D}, and 10 class II discs from the old Upper Scorpius (UpperSco) region \citep{2025ApJ...989....4A}. In this programme, the gas disc mass was found to decrease rapidly over time \citep{2025ApJ...989...10T, 2025ApJ...989....1Z}. Additionally, the radius of the gas discs was constrained only for the Lupus and UpperSco discs \citep{2025ApJ...989...10T}, with older discs exhibiting larger radii \citep{2025ApJ...989....1Z}. When comparing results on gas with those from the dust, the gas-to-dust mass ratio calculated in \cite{2025ApJ...989...10T} suggests a different timescale in the evolution of the two disc's components. The results on dust emission are consistent with the presence of dust traps in the discs of the sample. However, the AGE-PRO constraints on the dust constituent were limited: observations of dust emission at 0.15'' resolution were done at a single wavelength (ALMA Band 6, 1.3 mm), providing only a preliminary estimate on dust mass, dust disc size, and gas-to-dust mass ratio.

In this work, we present new ALMA Bands 3 and 5 continuum emission observations (3.1 and 1.5 mm, respectively) of the complete AGE-PRO survey. We analysed these new data by fitting the visibilities directly, and compare our results with those from AGE-PRO that used only Band 6 (1.3 mm) observations. By taking advantage of the AGE-PRO survey design to study disc evolution, with this multi-wavelength dataset we were able to infer dust properties and relate them to their evolution. 

This paper is organized as follows: Section \ref{sec:observations} presents the observations and provides information about the calibration and imaging process. Section \ref{sec:visfit} explains the visibility fitting and presents the direct results from this process. Section \ref{sec:fit} describes the multi-wavelength fitting to the radial profiles and shows the results for the inferred dust properties in this work. Section \ref{sec:discussion} discusses the main results and put them in context of evolution. Finally, section \ref{sec:conclusion} summarizes the main conclusions of this work.


\begin{figure*}
	\includegraphics[width=1.0\textwidth]{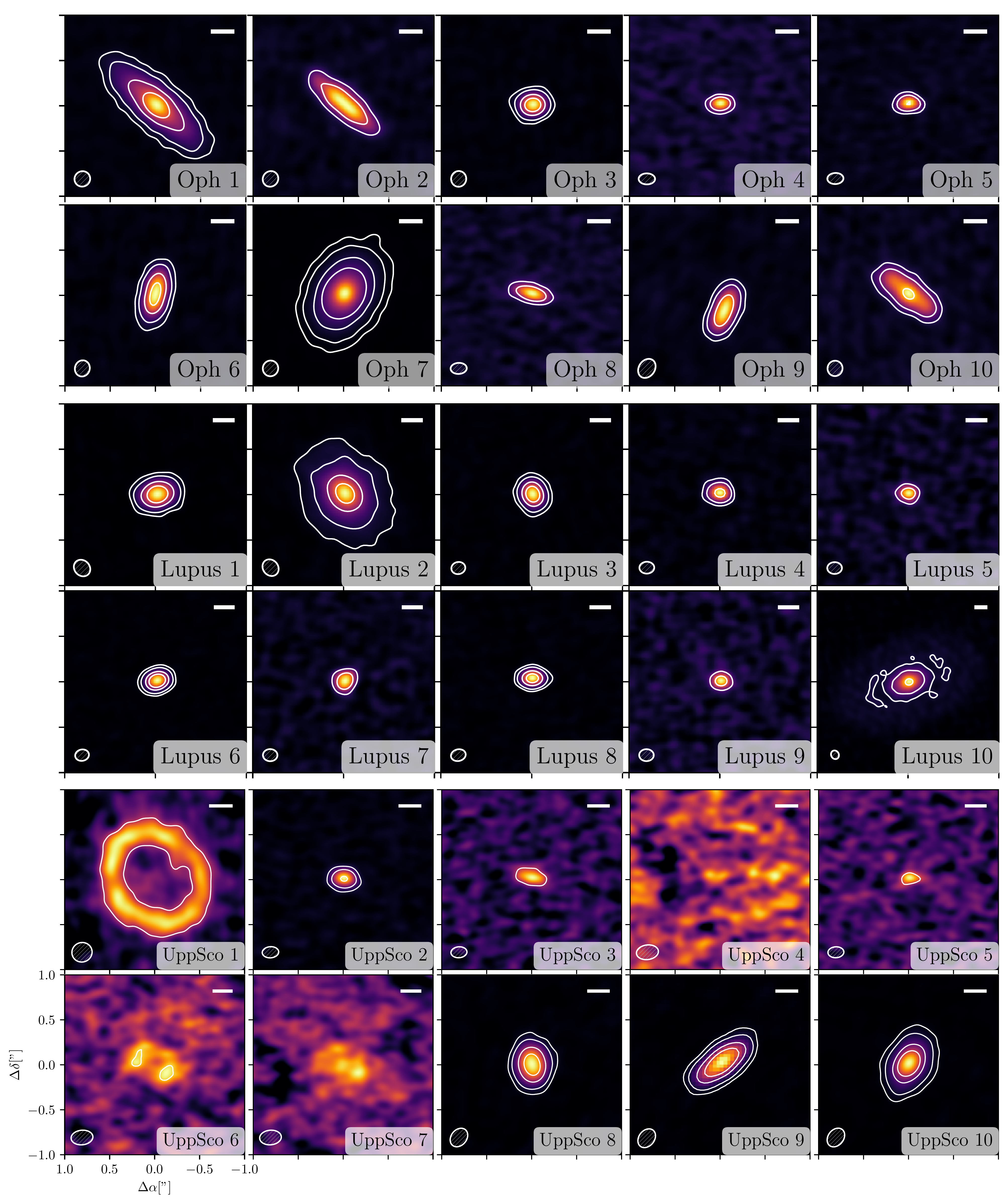}
    \caption{Band 5 (1.5 mm) continuum images of the whole AGE-PRO sample. The ellipse in the bottom left corner of each image marks the beam of the image. The total field of view (FOV) is of $2''$ for all the images except for Lupus 10, where the FOV  is of $4''$. The horizontal line in the top right of each panel correspond to a 30 au scale. The contours mark the $5\sigma$, $15\sigma$, $50\sigma$, and $95\sigma$ detections. The used nomenclature correspond to those adopted in the AGE-PRO Large Program and can be found in \protect\cite{2025ApJ...989....1Z}.}
    \label{fig:galery}
\end{figure*}

\section{Observations}\label{sec:observations}

We observed with ALMA Bands 3 and 5 (3.1 and 1.5 mm, respectively) the dust continuum emission of the 30 protoplanetary discs in the AGE-PRO sample, defined in \cite{2025ApJ...989....1Z}. The complete sample of 1.5 mm observations, and some of the 3.1 mm observations, were taken during Cycle 9 (project ID \texttt{2022.1.00638.S}; PI: L. P\'erez) with configuration C43-6 and C43-7 for the 1.5 and 3.1 mm observations, respectively, reaching resolutions between 0.15'' and 0.3''. For some discs in the sample, we used Band 3 archival data from project \texttt{2019.1.01111.S} (PI: A. Ribas) for Ophiuchus 1, 2, 3, 6, and 10, and from project \texttt{2015.1.00761.S} (PI: P. Sheehan) for Ophiuchus 7. In this paper we are using the AGE-PRO nomenclature, with discs numbered from 1 to 10 in each star forming region \citep[see][]{2025ApJ...989....1Z}.

We first calibrated the data using the ALMA pipeline within CASA \citep[Common Astronomy Software Applications,][]{McMullin2007}. In sources where the signal-to-noise ratio (SNR) was high enough ($20 < $ SNR) we applied additional self-calibration in CASA to improve the image quality and SNR. This process was performed using the NRAO's automated self-calibration script \citep{autoselfcal} \footnote{\url{https://science.nrao.edu/srdp/self-calibration-preview}}, which allows for rapid application of multiple self-calibration rounds in datasets that are obtained in a single ALMA configuration. For discs observed in multiple execution blocks (Ophiuchus 4, 5, UppSco 9, 10 in Band 3) the data were concatenated into a single measurement set prior to self-calibration. To do this, we first identified the centre of emission by fitting a Gaussian in the image plane, and then used the CASA tasks \texttt{fixvis} and \texttt{fixplanets} to adjust the phase and pointing centres accordingly. 

The automated self-calibration code employed defines imaging parameters, such as the mask, and the different solution intervals (\texttt{solint}) for gain calibration. The code attempts first phase self-calibration, and after stopping (either because of a decrease in SNR or a minimum \texttt{solint} reached), it tries amplitude self-calibration. For the images during self-calibration, we selected an intermediate Briggs weighting parameter (\texttt{robust = 0.5}) for the whole sample, except for faint sources in UpperSco (UpperSco 1, 4, 6, 7), for which we used \texttt{robust = 2} to reach the highest possible signal-to-noise ratio, at the expense of angular resolution.
Self-calibration was not successful in the whole sample, particularly for sources with low SNR (Ophiuchus 2, 4, 8, 9, Lupus 4, 5, 7, 9, and UpperSco 1, 3, 4, 5, 6, 7). For those, we used the data as calibrated only by the ALMA pipeline. Almost all sources were detected in both bands with the exception of UpperSco 4, which is the faintest source (0.13 mJy at 1.3 mm) in the entire survey and was not detected at either 1.5 or 3.1 mm (see Figures \ref{fig:galery} and \ref{fig:images_B3}), so we discard this source from the analysis in this paper.

\begin{figure*}
	\includegraphics[width=0.92\textwidth]{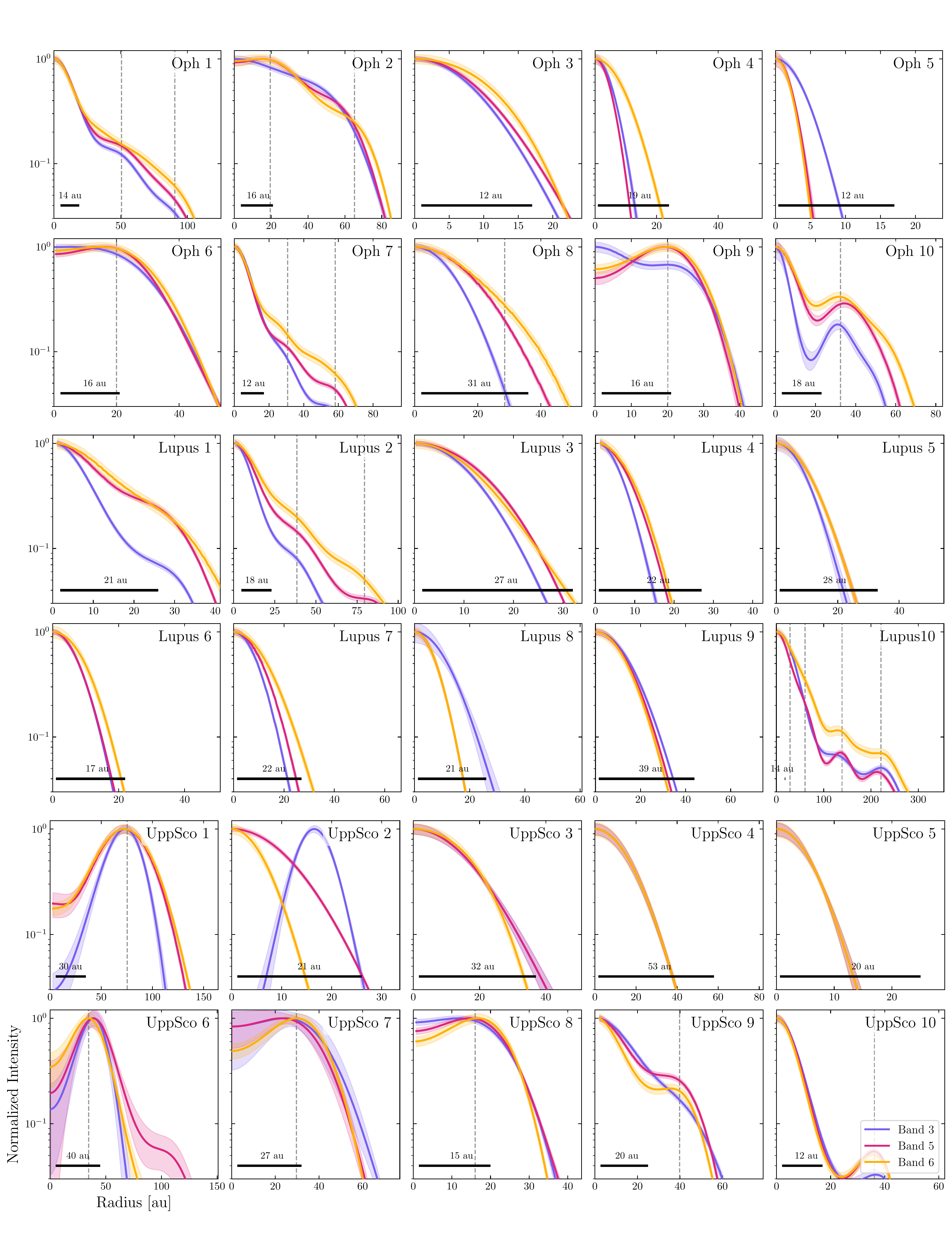}
    \caption{Normalized radial brightness profiles up to 100 au (300 au for Lupus 10) for the whole AGE-PRO survey in Bands 3, 5, and 6, the latter from \protect\cite{2025ApJ...989....9V}. Dashed vertical lines represent the detected bumps/ring-like structures locations in the discs from \protect\cite{2025ApJ...989....9V}. For each disc, the three radial profiles are convolved to the larger spatial resolution between the three observed wavelengths, which is represented by the horizontal line in the bottom of each panel. }
    \label{fig:radprofiles}
\end{figure*}


\begin{figure*}
	\includegraphics[width=0.9\textwidth]{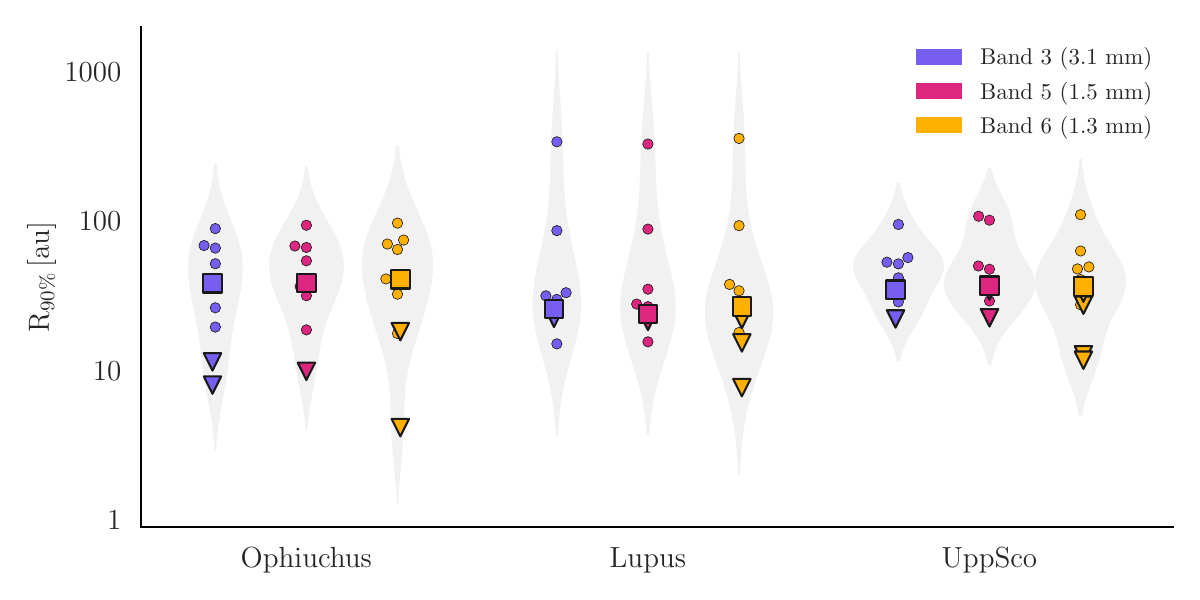}
    \caption{Estimated dust disc sizes ($R_{90\%}$) derived from radial brightness profiles at 3.1 mm (purple), 1.5 mm (pink), and 1.3 mm (yellow, from \citet{2025ApJ...989....9V}). Dots indicate the individual measurement per disc, while the square represents the median value per region. Triangles mark the disc size upper limit, for those discs in which the intrinsic resolution is larger than the measured disc size.}
    \label{fig:r90}
\end{figure*}

\section{Visibility fitting}\label{sec:visfit}

In order to recover sub-beam substructures in our sample of protoplanetary discs, we used the \texttt{Frankenstein (Frank)} code \citep{jennings2020}, which non-parametrically fits the observed interferometric visibilities. \texttt{Frank} uses a fast Gaussian process to reconstruct the radial brightness profiles $I_{\nu}\left(r\right)$ of a disc assuming azimuthal symmetry. 

In this work, we fit the visibility data of the new ALMA 3.1 and 1.5 \,mm observations using the same geometry (inclination and position angle) derived with Galario modelling of the data visibilities, and $R_{\text{out}}$ and $N$ values adopted in \cite{2025ApJ...989....9V} for the fit to the 1.3\,mm visibilities. For the \texttt{Frank}'s hyperparameters $\alpha_\texttt{Frank}$ and $w_{\text{smooth}}$, we tested different combinations to find a suitable fit to the visibilities. We observed that for discs with high signal-to-noise ratio, the impact of these hyperparameters is minimal on the inferred radial profile. However, this is not the case for discs with low SNR, where high-amplitude oscillatory artifacts can appear in the inferred brightness profile depending on the different combinations of hyperparameters. Thus, we tested 24 combinations for $\alpha_\texttt{Frank} \in [1.05, 1.1, 1.15, 1.2, 1.25, 1.3]$ and $w_{\text{smooth}} \in [10^{-4}, 10^{-3}, 10^{-2}, 10^{-1}]$, as these values are within the range suggested by \cite{jennings2020}. The radial brightness profile for each disc was chosen by visual inspection of the visibility fitting, selecting those with the minimal $\chi^2$ between model and data, and that also converge to zero brightness at largest deprojected uv distances, following the same methodology as in \cite{2025ApJ...989....9V} for the 1.3\,mm visibilities. In the case where the solutions from \texttt{Frank} were similar regardless of the hyperparameters used, we opted for the more conservative pair (higher $\alpha_\texttt{Frank}$ and $w_{\text{smooth}}$). For discs with low SNR, we consider the 1.3\,mm radial profile obtained by \cite{2025ApJ...989....9V} as a reference to exclude solutions that appear to be over-fitting and introducing oscillatory artifacts in the profiles, as the ones obtained for Ophiuchus 9 and UpperSco 2. The hyperparameters used for each disc are shown in Tables \ref{tab:oph_frank_params}, \ref{tab:lup_frank_params}, and \ref{tab:upp_frank_params}. 

Uncertainties in our inferred radial brightness profiles were estimated by considering two associated sources of uncertainty: the intrinsic error from the \texttt{Frank} fit ($\Delta I_{\nu_{\texttt{Frank}}}$), and the absolute flux calibration error corresponding to each ALMA band ($\Delta I_{\nu_{\text{calibration}}}$). Following the ALMA Technical Handbook \citep{cortes_2025_14933753}, we adopt a 5\% flux uncertainty for Bands 3 and 5, and 10\% for Band 6. Then, the final uncertainty at each radius was computed as

\begin{equation}
    \Delta I_{\nu} = \sqrt{\Delta I_{\nu_{\texttt{Frank}}}^{2} + \Delta I_{\nu_{\text{calibration}}}^{2}}.
\end{equation}

The radial brightness profiles for each disc in the sample and their associated uncertainties, are presented in Figure \ref{fig:radprofiles}, for the new 1.5 and 3.1 mm observations, and for comparison, the 1.3 mm radial profiles from \cite{2025ApJ...989....9V}. Note that $\Delta I_{\nu_{\text{calibration}}}$ dominates because $\Delta I_{\nu_{\texttt{Frank}}}$ is much smaller than any percentage of the flux calibration. Thus the total error in Figure \ref{fig:radprofiles} seems constant with radius. For all discs, the three profiles were convolved to the same resolution, which corresponds to the largest intrinsic \texttt{Frank} resolution among the three bands. This intrinsic \texttt{Frank} resolution is estimated by fitting the visibilities of a delta ring using the same uv-coverage and SNR as each observation, and by analysing the resulting point spread function (PSF), following \cite{sierra2024}. The resolution for each profile, in the three bands is shown in Tables \ref{tab:oph_frank_params}, \ref{tab:lup_frank_params}, and \ref{tab:upp_frank_params}.

\subsection{Dust disc sizes}

The radial profiles of protoplanetary discs do not exhibit a sharp outer edge or cut-off (see Figure \ref{fig:radprofiles}). Instead, the emission generally follows a decreasing trend that resembles a power-law distribution. This gradual decline makes it challenging to define a precise disc radius \citep[e.g.,][]{2023ApJ...954...41T}, as there is no obvious boundary separating the disc's emission from background noise. Consequently, establishing a consistent and physically meaningful definition for the disc size is essential for comparing different sources and datasets. 

Following the approach proposed by \cite{Tripathi2017}, we adopt a definition based on the integrated flux: we define the disc radius as the distance from the centre that encloses a defined fraction of the total continuum emission. Specifically, we use the radius that encloses 90\% of the total flux ($R_{90\%}$), consistent with the methodology employed by AGE-PRO at 1.3 mm \citep{2025ApJ...989....4A,2025ApJ...989....3D,2025ApJ...989....2R,2025ApJ...989....9V}. Additionally, we compute the radius enclosing 68\% of the total flux ($R_{68\%}$), as this has often been identified as a better proxy for protoplanetary discs sizes observed with low sensitivity \citep{miotello2022}, allowing us to compare the compactness of emission across discs and explore the relation between these two measurements. A summary of the dust disc sizes in the three studied wavelengths is shown in Figure \ref{fig:r90}, grouped by region. The same trend is found when comparing $R_{90\%}$ and $R_{68\%}$: there is a good consistency between the measurements for the same disc in the three bands, and we only see large deviations for unresolved discs (marked as triangles in Figure \ref{fig:r90}), with sizes smaller than about 20 au (smaller than the intrinsic resolution of the radial profile obtained with \texttt{Frank}, see Figure \ref{fig:radprofiles}). 
The measured $R_{90\%}$ for each disc in the sample in the three studied wavelengths are reported in Table \ref{tab:properties}. Note that some of the discs have a smaller extent than the profile resolution (as is the case of Ophiuchus 4 and 5, Lupus 4, 5, 8, and 9, and UpperSco 2, 3, 4, and 5), we are going to refer to this discs as unresolved in the visibility plane.

\subsection{Spectral indices}\label{sec:spectral_index}

The spectral index, $\alpha$, is defined as the exponent of the spectral energy distribution in the millimeter range, this is $F_{\nu} \propto \nu^{\alpha}$. Neglecting effects of scattering, the emitted intensity is given by

\begin{equation}\label{eq:intensity}
    I_{\nu} = B_{\nu}(T)(1-\text{exp}(-\tau_{\nu})),
\end{equation}

\noindent where $B_{\nu}(T)$ is the Planck function at temperature $T$ and $\tau_{\nu} = \Sigma_{d}\kappa_{\nu}/\cos i$ is the absorption optical depth, with $\kappa_{\nu}$ the absorption coefficient and $i$ the disc inclination. If the dust opacity $\kappa_{\nu}$ follows a power-law ($\kappa \propto \nu^{\beta}$), the frequency dependency of $I_{\nu}$ in the Rayleigh-Jeans limit becomes $\alpha = 2 + \beta$ when the emission is optically thin ($\tau_{\nu} \ll 1$), and $\alpha = 2$ for optically thick emission ($\tau_{\nu} \gg 1$). Spectral indices are typically measured over a specific wavelength range, between frequencies $\nu_{1}$ and $\nu_{2}$, given by

\begin{equation}\label{eqalpha_global}
    \alpha = \frac{\text{ln}F_{\nu_1} - \text{ln}F_{\nu_2}}{\text{ln}\nu_1 - \text{ln}\nu_2},
\end{equation} 

\noindent where $F_{\nu_{1}}$ and $F_{\nu_{2}}$ are the fluxes measured at frequencies $\nu_{1}$ and $\nu_{2}$ respectively. The uncertainties in $\alpha$ are estimated following the procedure in \cite{chiang2012}. The value of $\alpha$ is consistent within the uncertainties when computed from the total flux of the \texttt{Frank} fits, and when computed from the total flux from the image. 

\begin{figure}
	\includegraphics[width=0.9\columnwidth]{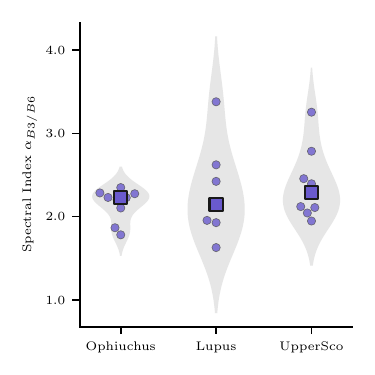}
    \caption{Spectral index calculated between 1.3 and 3.1 mm emission, separated by region. The dots indicate the individual measurements and the squares mark the median value per region. }
    \label{fig:alpha}
\end{figure}

The spectral index of dust in the interstellar medium (ISM) has typical values of $\alpha \sim 3.7$ \citep{draine2006}. If $\alpha$ is below this value, we can assume that grain growth is taking place in the disc \citep[e.g.,][]{perez2012}. If $\alpha \leq$ 2, it may indicate that the disc emission is optically thick with self-scattering or important temperature gradients along the line of sight \citep{sierra+lizano2020}. The integrated spectral indices between Bands 3 and 6, and their associated uncertainties for each disc in our sample are reported in Table \ref{tab:properties} and summarized in Figure \ref{fig:alpha}. We observe a consistency of the spectral index across the three star-forming regions, with median values near $\alpha \sim 2$ but the spread clearly increasing from the Ophiuchus discs to the older discs.

\section{Constraints on Dust Properties}\label{sec:fit}

The spatially integrated dust continuum emission at millimetre wavelengths has been used to infer dust properties of protoplanetary discs, including dust temperature, optical depth, maximum grain size, and dust disc mass. Owing to the lack of multi-wavelength observations at high angular resolution, the spatial modelling of spectral energy distributions (SED) has only been done for a small fraction of discs \citep[e.g.][]{perez2012, perez2015, 2016A&A...588A..53T, carrasco2019, macias2021, sierra2025, carvalho2024, guerra-alvarado2024}. In this work, we were able to spatially resolve 20 of the dust discs in the AGE-PRO sample, in which we modelled the SED at sub-millimetre wavelengths as a function of radius. For the unresolved discs, we adopted the typical methodology and fit the spatially integrated SED. \\

For the 20 resolved discs, the SED is fitted at each radius (i.e., the three radial profiles) using the intensity of the dust continuum emission at a frequency $\nu$, including both absorption and scattering as in \cite{sierra+lizano2020}, which is given by 

\begin{equation}\label{eq5}
    I_{\nu} = B_{\nu}(T_{d})[1-\text{exp}(-\tau_{\nu}/\mu) + \omega_{\nu}F(\tau_{\nu}, \omega_{\nu})],
\end{equation}

\noindent where $B_{\nu}$ is the Planck function evaluated at the dust temperature $T_{d}$, $\tau_{\nu}$ is the total optical depth, $\mu = \cos(i)$ with $i$ the inclination of the disc, $\omega_{\nu}$ is the dust albedo, and $F(\tau_{\nu}, \omega_{\nu})$ is the contribution of scattering to the emergent intensity. The main effect of scattering is to decrease the emergent intensity near the optically thick regime, making it possible to explain spectral indices below 2 \citep[see, e.g.,][]{sierra+lizano2020}. Dust grains are assumed to follow a particle size distribution (number of grains per size bin) given by $n(a)da \propto a^{-p}da$, where the minimum grain size is fixed to 0.05 $\mu$m, and the maximum grain size $a_{max}$ is a free parameter. To account for grain growth and settling in the disc interior \citep[e.g.,][]{dullemond2004}, we assume a particle size distribution slope of $p = 3.0$. The expected value  of $p$ from simulations is between 2.5 and 3.5, when the maximum grain size is regulated by drift and fragmentation, respectively \citep{birnstiel2012}.  \cite{sierra2024} tested $p$ values within this range, and found that the qualitative conclusions on dust properties are not too sensitive to this particular choice. 

For this work we used different opacities to fit the spatially-resolved multi-wavelength observations of the discs: the default DSHARP opacities \citep{birnstiel2018} and Ricci's \citep{ricci2010} opacities. In both cases, we consider spherical grains and use the DSHARP opacity tools \citep{birnstiel2018}. The main difference between these two models lies in the inclusion of carbon grains in the Ricci opacities, which at millimetre wavelengths increases the absorption and reduces the albedo. This leads to less massive discs when inferring dust surface density from multi-wavelength observations, or increases the total intensity for fixed dust surface density locations when using the Ricci's opacities. \citep{sierra2024}.

Equation \ref{eq5} can be evaluated given the dust temperature $T_{\text{d}}$, maximum grain size $a_{\text{max}}$ (which controls $\omega_\nu$ and $\kappa_\nu$), and dust surface density (directly related with the optical depth as $\tau_\nu  = \kappa_\nu \Sigma_{\text{d}}/(1-\omega_\nu)$). We followed the same procedure as in \cite{sierra2024} and used a Markov Chain Monte Carlo (MCMC) method using 24 walkers and 5000 steps to explore, at each radius, the space parameter of $T_{\text{d}}$, $a_{\text{max}}$, and $\Sigma_{\text{d}}$. The $\chi^2$ of the posterior distribution is defined as 

\begin{equation}\label{eq:chi2}
    \chi^2 = \sum_n \left( \frac{I_{\nu_n} - I_{\nu_n}^{\mathrm{model}}}{\Delta I_{\nu_n}}\right)^2
\end{equation}

\noindent that is, the probability of a model with temperature $T_d$ decreases if it deviates from $T_{\text{prior}}$. The latter is computed from the temperature of a passively irradiated disc at radius $r$ and the stellar luminosity $L_*$, $T_{\text{prior}} = (0.02L_{*}/8\pi\sigma_{B}r^{2})^{1/4}$. We use a wide temperature width $\sigma_{T_{\text{prior}}} = 50$ K to ensure that this does not dominate the posterior probability distribution. The inclusion of a temperature prior decreases the degeneracy between the free parameters, especially when the number of wavelengths used to fit the SED is less than or equal to the number of free parameters. 

\begin{figure}
	\includegraphics[width=\columnwidth]{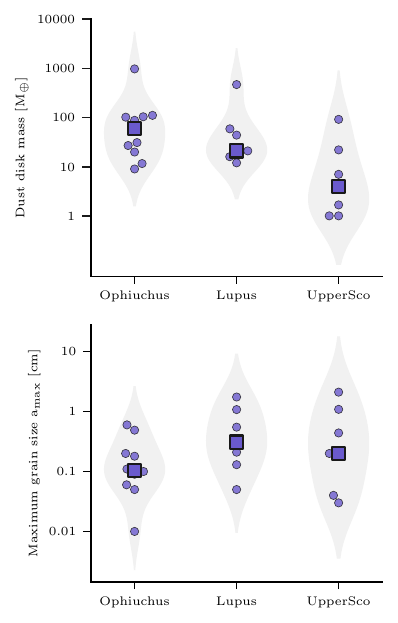}
    \caption{Dust disc masses (\textit{top panel}), and radially averaged maximum grain size (\textit{bottom panel}) of the AGE-PRO discs obtained assuming the Ricci's opacities \protect\citep{ricci2010}. The square represent the median value of each property in each region.}
    \label{fig:ricci}
\end{figure}

The explored parameter space is log-spaced between 300 $\mu$m $\leq a_{max} \leq$ 30 cm for the maximum grain size, $10^{-6}$ g cm$^{-2} \leq \Sigma_d \leq$ 10 g cm$^{-2}$ for the surface density of solids, and 3 K $\leq T_d \leq $ 300 K for the dust temperature. A summary of the dust disc masses derived directly from the dust surface density obtaining with this modelling, and the radially averaged maximum grain sizes, obtained when adopting the Ricci's opacities is showed in Figure \ref{fig:ricci} and on Table \ref{tab:properties}. The results on maximum grain size using DSHARP opacities are shown in Figure \ref{fig:amax}.

\begin{table*}
\centering
\caption{Dust disk sizes ($R_{90\%}$) measured from \texttt{Frank} radial profiles, and millimetre fluxes and spectral indices (B3/B6) derived from the images for three representative discs in the sample. Results from 1.3 mm observations, reported in \protect\cite{2025ApJ...989....4A, 2025ApJ...989....2R, 2025ApJ...989....3D, 2025ApJ...989....9V}, are shown for comparison. The reported dust mass, maximum grain size, and gas-to-dust ratio correspond to that obtained with the Ricci's opacities \protect\citep{ricci2010}. A complete version of this table which include also columns with $R_{68\%}$, the synthesized beam, rms noise and SNR for the Bands 3 and 5 images, as well as the dust properties (maximum grain size, dust disc mass, and gas-to-dust mass ratio) derived using Ricci's and DSHARP opacities, for the 30 discs in the AGE-PRO sample, is available in \href{https://doi.org/10.5281/zenodo.23535413}{https://doi.org/10.5281/zenodo.23535413}}

\label{tab:properties}
\begin{tabular}{@{}cccccccc@{}}
\hline
Source       & \begin{tabular}[c]{@{}c@{}}ALMA \\ Band\end{tabular} & \begin{tabular}[c]{@{}c@{}}R$_{90\%}$\\ {[}au{]}\end{tabular}             & \begin{tabular}[c]{@{}c@{}}Flux\\ {[}mJy{]}\end{tabular}  & \begin{tabular}[c]{@{}c@{}}$\alpha$\\ {[}B3/B6{]}\end{tabular} & \begin{tabular}[c]{@{}c@{}}Dust mass\\ {[}M$_{\oplus}${]}\end{tabular}  & \begin{tabular}[c]{@{}c@{}}$a_{\text{max}}$\\ {[}cm{]}\end{tabular} & \begin{tabular}[c]{@{}c@{}}gas-to-dust\\ {[}mass ratio{]}\end{tabular}    \\ \hline

Ophiuchus 2  & \begin{tabular}[c]{@{}c@{}}3\\ 5 \\6 \end{tabular} & \begin{tabular}[c]{@{}c@{}}66 $\pm$ 10\\ 66 $\pm$ 10 \\ 69 $\pm$ 10\end{tabular} & \begin{tabular}[c]{@{}c@{}}5.10 $\pm$ 0.26\\ 15.8 $\pm$ 0.8 \\ 27.5 $\pm$ 0.14\end{tabular} & 2.01 $\pm$ 0.13  & 110.7 $\pm$ 51.5 & 0.58 $\pm$ 0.1 & 22.7\\ \hline

Lupus 2      & \begin{tabular}[c]{@{}c@{}}3\\ 5\\ 6\end{tabular} & \begin{tabular}[c]{@{}c@{}}85.0 $\pm$ 0.5\\ 87.1 $\pm$ 0.7\\ 91.1 $\pm$ 1.4\end{tabular} & \begin{tabular}[c]{@{}c@{}}9.7 $\pm$ 0.5\\ 35.1 $\pm$ 1.7 \\ 76.4 $\pm$ 0.1\end{tabular} & 2.35 $\pm$ 0.12 & 55.1 $\pm$ 21.5 & 0.045 $\pm$ 0.001 & 69.0 \\ \hline

UpperSco 1   & \begin{tabular}[c]{@{}c@{}}3\\ 5 \\6\end{tabular} & \begin{tabular}[c]{@{}c@{}}83.4 $\pm$ 2.3\\ 94.1 $\pm$ 1.1\\87.64 $\pm$ 0.62 \end{tabular} & \begin{tabular}[c]{@{}c@{}}1.72 $\pm$ 0.09\\ 9.9 $\pm$ 0.46\\ 16.1 $\pm$ 0.1\end{tabular} & 2.66 $\pm$ 0.13 & 1.14 $\pm$ 0.19  & 0.032 $\pm$ 0.001 & 1077 \\

\hline
\end{tabular}
\end{table*}


\section{Discussion}\label{sec:discussion}

\subsection{Evolution of dust disc sizes}\label{sizes_evol}
From the radial profiles obtained with \texttt{Frank}, we measure the dust disc sizes as the radii that encloses 90\% of the total flux emission, reported in Table \ref{tab:properties}. As shown in Figure \ref{fig:r90}, the constrained sizes are consistent at the three wavelengths, and there are no large deviations between the multi-wavelength measurements for spatially-resolved discs. However, differences in disc sizes at different wavelengths appear for discs that are marginally resolved or unresolved, as those disc sizes cannot be well constrained. 

We compile the survey results on dust disc sizes ($R_{90\%}$), in violin plots presented in Figure \ref{fig:r90}, where the x-axis can be understood as time increasing to the right: discs in Ophiuchus are the youngest (with ages below 1 Myr), discs in Lupus are middle-age, ages between 1-3 Myr, and UpperSco discs being the oldest of the survey, with ages larger than 3 Myr. The dust disc size trend reported by AGE-PRO \citep{2025ApJ...989....9V} with only 1.3\,mm observations is recovered in our new 1.5 and 3.1\,mm observations. There appears to be no clear evolutionary trend, as sizes are similar in the whole survey. Lupus discs have a slightly smaller median dust disc sizes than Ophiuchus and UpperSco discs. 

A constant value of the dust disc size, across wavelengths and age, contrast our theoretical understanding of the growth and evolution of solids in a disc. \cite{weidenschilling1977} established that, in a smooth disc, larger particles should disappear due to radial drift in timescales much shorter than the disc lifetime. Thus, as discs age, their size, traced by the extent of continuum dust emission, should decrease if radial drift is fully acting. Additionally, differences in disc size between the three observed wavelengths are expected, since longer wavelengths trace larger grains that are more affected by radial drift. However, we are not observing these effects in our AGE-PRO sample. The lack of evolutionary trend in this property can be a selection bias, since the UpperSco discs in the sample are survivors \citep{ping2026}, and thus the discs that may be born as the brightest and biggest in the region, nevertheless, \cite{2025MNRAS.543.2723P} compared a more complete sample of Lupus and UpperSco discs and found no significant difference on the $R_{90\%}$. Therefore, this result may be implying that a physical mechanism is halting the inward radial drift of particles. This result may be an indicator of dust trapping in these discs, as particles preferentially drift toward higher gas pressure and their drift can be reduced, or stopped, by localized gas pressure maxima \citep{pinilla2012}. For example, \cite{2020A&A...635A.105P, 2020ApJ...898...36L} showed that the presence of pressure bumps in discs can lead to the absence of significant differences in the dust extent among wavelengths.
In particular, as part of the single-wavelength AGE-PRO analysis, \cite{2025ApJ...989....6K} ran dust evolution models for the disc population traced by AGE-PRO, finding that the millimetre continuum extent is often controlled by the location of the outermost dust trap, even at early ages ($<1$ Myr). 

In addition to the lack of evolutionary trend for dust disc sizes, the large amount of substructured discs detected in the three bands, and the dust surface density and maximum grain size profiles obtained through the radial multi-wavelength fitting are consistent with the presence of dust traps in some discs of the sample. In general, we detect narrower ring-like substructures in the radial profiles of long wavelength observations (see Figure \ref{fig:radprofiles}), which is expected when dust size differential trapping occurs \citep{pinilla2015}. Furthermore, our multi-wavelength modelling revealed local maxima of dust surface density and maximum grain size at the location of the detected rings in Ophiuchus 10, Lupus 10, and UpperSco 1 (see Figures \ref{fig:ophiuchus}, \ref{fig:lupus}, and \ref{fig:uppsco}). For other substructured discs, there is no correlation between the location of the rings and the peaks in the profiles of $\Sigma_\text{d}$ or $a_{\text{max}}$. Higher resolution and sensitivity observations at multiple wavelengths can help disentangle the presence of dust traps on those discs.

\subsection{Spatially-integrated spectral indices}

As explained in Section \ref{sec:spectral_index}, differences in the spectral index $\alpha$, between individual discs or among regions, directly reflect differences in dust properties. 
A summary of the global spectral indices per region is shown in Figure \ref{fig:alpha}. The computed spectral index $\alpha_{\text{1.3mm/3.1mm}}$ shows the lowest median in the Lupus region and the largest for UpperSco discs. The AGE-PRO collaboration was able to calculate spectral indices from 1.3 and 0.9mm observations for Lupus and UpperSco discs \citep{2025ApJ...989....1Z}, their results present the same behaviour as the one obtained here for longer wavelengths. 

In our sample, we obtain $\alpha < 2$ for three resolved discs (Ophiuchus 3, Lupus 6 and 7) and three unresolved discs (Lupus 4, 8, 9). For the resolved discs our analysis shows that these are heavily optically thick ($\tau \gg 1$ over the entire or a large fraction of the disc extent, even at Band 3 observations, see Fig. \ref{fig:ophiuchus}). For the unresolved discs, we compute the optically thick fraction defined in \cite{tazzari2021}. The optically thick fraction is the relation between the observed flux and the flux of a completely optically thick source with the same radius ($\rho_{x}$) as the disc. This is given by 

\begin{equation}
    \mathcal{F} = \frac{x L_{\rm mm}}{2\pi \int_{0}^{\rho_x}B_{\nu}(T_{\rm d})\rho'd\rho'},
\end{equation}

\noindent where $x$ is the fraction of the emission chosen for measuring the disc size, e.g. 0.9 for $R_{90\%}$, and $L_{\rm mm}$ is the disc luminosity at millimetre wavelengths. Since this value depends on the size of the disc, it is difficult to constrain the fraction for unresolved discs. However, we estimated it by adopting the upper limits on disc size from \cite{2025ApJ...989....9V} in the case of Lupus 4 and 8, while we adopt the $R_{90\%}$ value found in the high-resolution observations of \cite{guerra2025} for Lupus 9. The estimated optically thick fractions indicate that emission is optically thick across all three observed bands, with values of 0.9 for Lupus 4 and Lupus 8, and 1.0 for Lupus 9. Therefore, all the spectral indices below 2 in the AGE-PRO sample can be explained by optically thick emission in the presence of self-scattering or temperature gradients. Since scattering decreases the intensity relative to pure black-body emission, $\mathcal{F} \sim 1$ could be explained by optically thick emission from low-albedo grains (albedo $\lesssim 0.4$; Figure 1 of \cite{sierra+lizano2020}). However, this regime is inconsistent with the observed $\alpha < 2$, as significant scattering is required to explain such low spectral indices. Therefore, temperature gradients traced at different wavelengths provide a more likely explanation for our dataset.

\begin{figure}
	\includegraphics[width=\columnwidth]{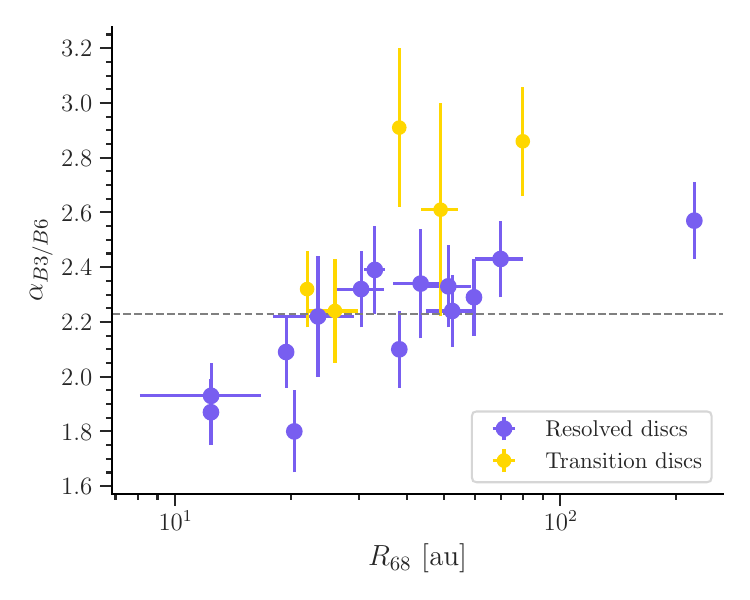}
    \caption{Spatially-integrated spectral indices ($\alpha_{\text{1.3mm/3.1mm}}$) vs. dust disc sizes ($R_{68\%}$) for the resolved sources in our sample. The yellow dots represent the measured spectral index of the detected transition discs in the survey, while the purple ones are those without an inner cavity. The horizontal line marks the median spectral index $\alpha_{\text{0.87mm/3.1mm}}$ found for Lupus in \protect\cite{tazzari2021}.}
    \label{fig:alpha_vs_r}
\end{figure}

\cite{tazzari2021} studied 26 bright discs in the Lupus region, finding a relation between spectral indices and dust disc sizes, finding that higher $\alpha$ values are observed in larger discs. Figure \ref{fig:alpha_vs_r} shows a comparison between the spectral indices and dust disc sizes ($R_{68\%}$) for the resolved sources in our sample. We did not include unresolved sources since their sizes are not well constrained. In our sample, transition discs (discs with resolved cavities, defined in Table 2 of \cite{2025ApJ...989....9V}) appear to have larger spectral indices when compared to discs of similar sizes. \cite{pinilla2014} showed that this trend arises because of the absence of dust traps in the inner cavity. Without dust traps, dust in the inner region of the disc could be removed due to radial drift, which is known to affect larger grains more. Thus, smaller grains are more likely to stay in the cavity region of these discs, increasing the value of the spatially-integrated spectral index. 

\begin{figure}
	\includegraphics[width=\columnwidth]{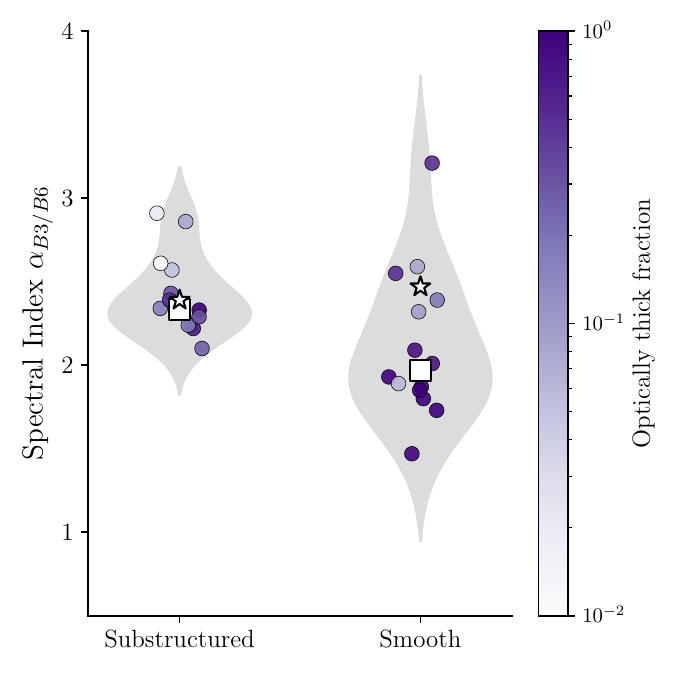}
    \caption{Spectral indices between 1.3 and 3.1 mm observations, separated by discs with substructures (on the left), and discs with no detected substructures at the observed resolution (on the right). Colours indicate the optically thick fraction of each disc as defined in \citet{tazzari2021}. The white squares mark the median of the measured values for substructured and smooth discs, while the star is the median considering only optically thin discs (with fraction less than 0.6).}
    \label{fig:alpha_subs}
\end{figure}

The detection of disc substructures makes it natural to compare the spectral index in discs with and without substructures. Models predict higher spectral indices in completely smooth discs, since the presence of optically thick ring-like substructures leads to lower values of $\alpha$ \citep{2026MNRAS.tmp..391L}. However, our results in Fig.~\ref{fig:alpha_subs} show the opposite trend. One possible explanation for this difference, can be that the discs labelled as ``smooth'' in this sample are those without substructures larger than the achieved resolution \citep[see][]{2025ApJ...989....9V}, then it is possible that some of those discs could be hiding substructures and leading then to lower spectral indices. Alternatively, the difference between these observations and the models from \cite{2026MNRAS.tmp..391L} can be due to optical depth effects, since from our observations we obtain that discs are optically thick even in Band 3 (see Figures \ref{fig:ophiuchus}, \ref{fig:lupus}, and \ref{fig:uppsco}), while the models from \cite{2026MNRAS.tmp..391L} are optically thin at millimetre wavelengths, leading to higher spectral indices. Further, all discs with $\alpha < 2$ are classified as smooth discs at our resolution, and their emission is optically thick, leading to a lower median when compared to the resolved substructured discs, which present optically thinner emission. When considering only optically thin discs, we recover the same trend as predicted by dust evolution models presented in \cite{2026MNRAS.tmp..391L}.

\subsection{Evolution of dust disc masses}

Our multi-wavelength analysis provides direct constraints on the dust disc masses and the maximum grain sizes for each individual source. Analogue to the comparison of dust disc sizes presented in Sect.~\ref{sizes_evol}, these quantities can be examined across regions, and thus, across different ages. Such an approach allows us to trace how the dust content and grain size evolve over time, offering insights into the efficiency of grain growth and the possible depletion of solids in protoplanetary discs.

\begin{figure*}
	\includegraphics[width=\textwidth]{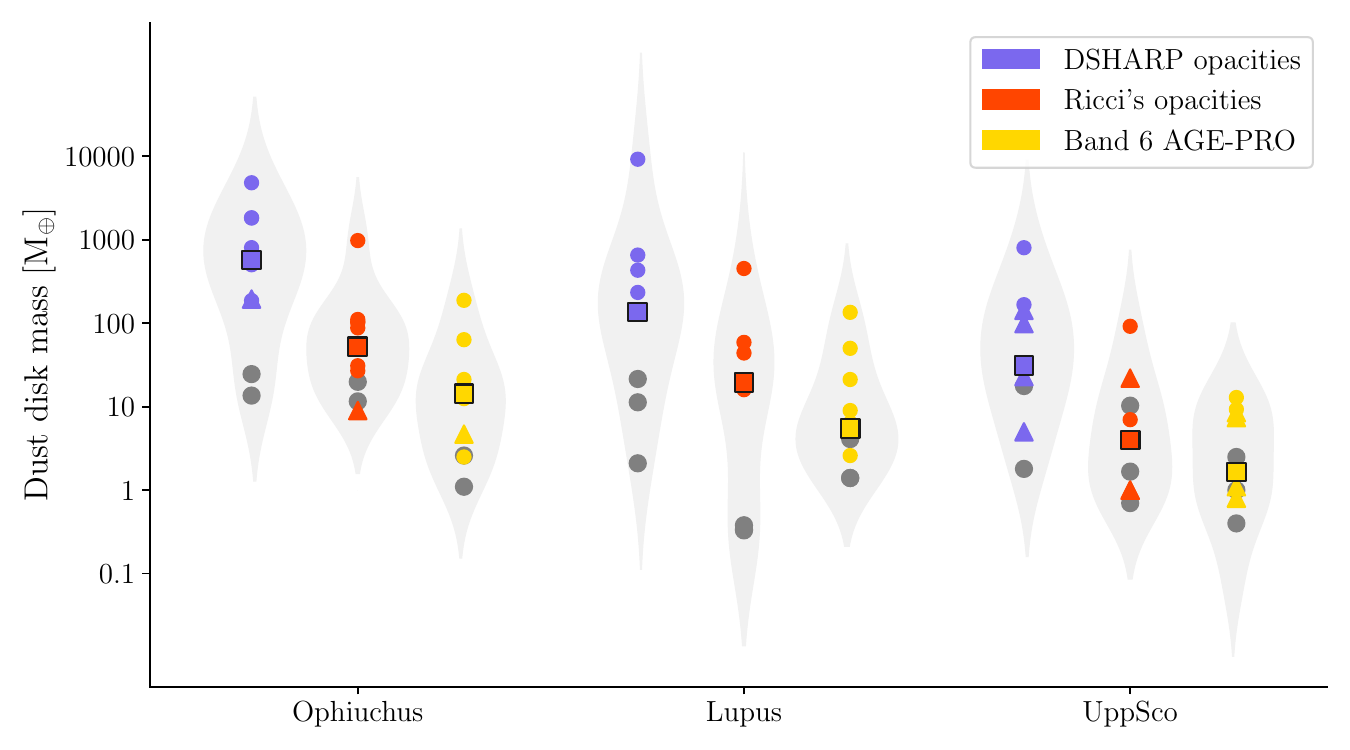}
    \caption{Dust disc masses per region obtained from multi-wavelength analysis adopting DSHARP (purple) and Ricci's (orange) opacities, and those obtained through the single wavelength analysis performed by AGE-PRO (yellow, from \protect\cite{2025ApJ...989....2R, 2025ApJ...989....3D, 2025ApJ...989....4A}). Grey points indicate dust masses for unresolved discs, while triangles mark the derived masses for transition discs.}
    \label{fig:mass}
\end{figure*}

Dust disc masses were estimated from the AGE-PRO single-wavelength observations at 1.3 mm in Band 6 \citep{2025ApJ...989....4A, 2025ApJ...989....3D, 2025ApJ...989....2R}, following the standard approach of assuming optically thin dust continuum emission and deriving dust disc masses from the equation: 

\begin{equation}\label{eq:mass_agepro}
	M_{\text{dust}} = \frac{d^2}{B_{\nu}(T_{d})\kappa_{\nu}}F_{\nu},
\end{equation}

\noindent where $d$ is the distance to the object, $B_{\nu}(T_\text{d})$ is the Planck spectrum at the average dust temperature $T_{\text{d}}$, and $\kappa_\nu$ is the dust opacity. The AGE-PRO collaboration adopted $T_{\text{d}} = 20$\,K, and $\kappa_\nu = 2.3 \left(\frac{\nu}{230 \textrm{GHz}} \right)^{\beta}$ cm$^2$ g$^{-1}$, with $\beta = 1$. For comparison, $\kappa_{1.3\text{mm}} = 5.56$ cm$^2$g$^{-1}$ for DSHARP opacities, and $\kappa_{1.3\text{mm}} = 18.16$ cm$^2$g$^{-1}$ for Ricci's opacities. In our approach, and by using multi-wavelength observations, we did not need to assume optically thin emission or a particular dust temperature. Figure \ref{fig:mass} presents the inferred dust mass of each disc obtained in this work. Although the dust opacity choice introduce large uncertainties to the derived dust disc masses, with either DSHARP or Ricci’s dust opacity, the same trend is observed when comparing the three regions, and thus, with age: the solid mass in discs decreases over time. 

Although these methodologies result in the same trend of dust mass decreasing over time, we note that single-wavelength analysis of this property differs by a factor of $\sim3-4$ and $\sim 20-40$ when compared to our multi-wavelength analysis employing Ricci’s and DSHARP opacities, respectively. As is seen in Figure \ref{fig:mass}, the choice of dust opacity significantly influences the derived dust properties, and in this case, the dust disc mass. Specifically, the difference in dust mass between DSHARP or Ricci opacities corresponds to a factor of $\sim$10 in the case of Ophiuchus, $\sim$9 for Lupus, and $\sim$4 for UpperSco, on average. The difference between regions might arise from the different optical depth on the discs: while UpperSco discs are less optically thick (see Figure \ref{fig:uppsco}), Ophiuchus and Lupus discs tend to be more optically thick at the three wavelengths, leading to higher dependency on the chosen opacities.
In general, although both opacities used in this work have similar compositions, with water ice and astronomical silicates, Ricci’s opacities include carbon grains, while DSHARP includes refractory organics. This moderate compositional difference results in higher values of opacity for Ricci’s than DSHARP, at a given wavelength and for the same grain size. This directly translates to less massive discs when adopting the dust composition of Ricci's to infer dust surface density with multi-wavelength fitting \citep{sierra2024}. 

As stated in \cite{viscardi2025}, the inclusion of optically thinner ALMA 3.1\,mm (Band 3) observations provides better constrains on dust disc masses. However, the optical depths derived in our modelling shows optically thick emission at the three wavelengths observed for most discs on the sample (in particular for Ophiuchus and Lupus discs), either adopting  Ricci's or DSHARP opacities (see Fig. \ref{fig:ophiuchus}, \ref{fig:lupus}, and \ref{fig:uppsco}), suggesting that in those discs we might be underestimating the dust disc mass \citep{viscardi2025}. 

\begin{figure*}
	\includegraphics[width=\textwidth]{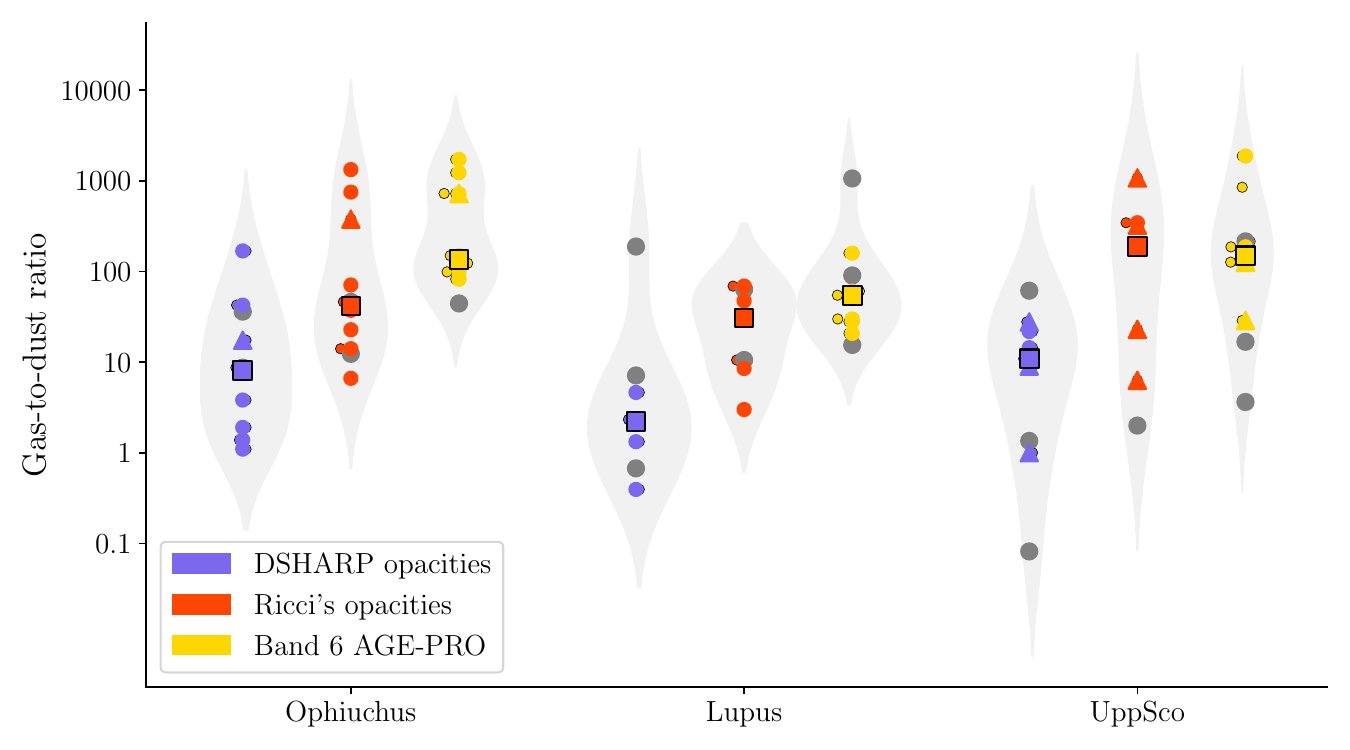}
    \caption{Gas-to dust mass ratio obtained using the multi-wavelength derived dust disc masses with DSHARP (purple) and Ricci's (orange) opacities. The values estimated in \protect\cite{trapman2025} from the 1.3\,mm only observations are shown in yellow for comparison.}
    \label{fig:g2dratio}
\end{figure*}

Nevertheless, we see a decline in dust mass from Ophiuchus (<1 Myr) through Lupus (1–3 Myr) and eventually UpperSco (>3 Myr), this can be explained by models that invoke leaky or weak pressure traps that keep the dust continuum extent constant as we observe (see Fig. \ref{fig:r90}), but also allow some drift to happen, reducing $M_{\text{dust}}$ over time \citep{2020A&A...635A.105P, 2025ApJ...989....6K, 2026MNRAS.tmp..391L}. As shown by \cite{birnstiel2012}, radial drift can deplete the outer disc of millimetre-emitting dust within just a few Myr. Concurrently, this pebble flux accumulates in the inner regions and promotes the formation of planetesimals and larger bodies, effectively locking up mass in objects too large to be effectively traced by (sub-)mm observations \citep{godines2025}. Observationally, a similar trend was observed in previous studies that traced dust mass at a single wavelength \citep[e.g.,][]{ansdell2016, pascucci2016}. The agreement between ALMA surveys and theoretical drift coagulation models confirms that, although large grains concentrate and grow into planetesimals, the observed total dust mass declines as discs age. We note that the differences between regions are similar to the uncertainties of the different opacities (a factor $\sim$10), therefore we cannot exclude the possibility of the opacities being different between regions or evolving over time, which would imply that the observed decreased does not reflect the actual dust disc masses depletion over age.

\subsection{Evolution of gas-to-dust ratio}

Understanding the gas-to-dust ratio in protoplanetary discs is essential for unveiling the processes that govern disc evolution and planet formation. Since most observational constraints rely on dust emission, the gas-to-dust ratio acts as a critical conversion factor for estimating the total disc mass. Moreover, it directly influences the efficiency of planet formation by controlling the availability of solid material for the growth of planetesimals and planetary cores \citep{testi2014, birnstiel2016, william+cieza2011}. Variations in the gas-to-dust ratio can therefore provide key insights into the initial conditions and evolutionary pathways in protoplanetary discs. 

Our better constraints on dust disc masses lead to better estimation of gas-to-dust mass ratios. Figure \ref{fig:g2dratio} shows our results directly obtained from using the dust disc masses from this work and the gas masses derived by the AGE-PRO collaboration in \cite{2025ApJ...989...10T}. Our results confirm those found by \cite{2025ApJ...989...10T}; the lowest median value of gas-to-dust ratio is obtained for the Lupus discs, while the largest is found for discs in the UpperSco region. 

The increase of gas-to-dust mass ratio in older discs is consistent with dust evolution models in which radial drift has removed large amounts of dust from the disc, either because dust traps are absent or because they are too inefficient to prevent most of that loss \citep{birnstiel2012, drazkowska2023}. However, this could also be interpreted as planet formation taking place in the oldest discs, where the dust traced at long wavelengths in younger discs has been aggregated into planetesimals and removed from the disc emission of older systems. Also, a combination of these effects may be taking place.

\begin{figure*}
	\includegraphics[width=\textwidth]{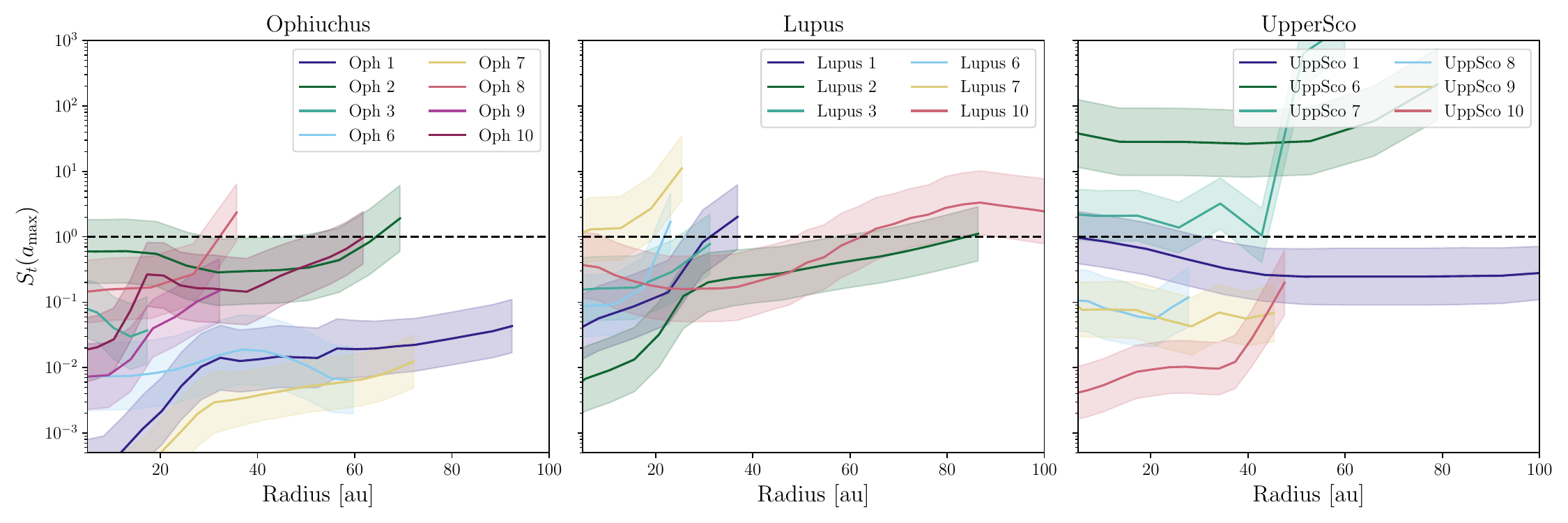}
    \caption{Stokes number ($S_t$) derived from dust properties obtained from our multi-wavelength analysis for the resolved discs in the AGE-PRO sample adopting Ricci's opacities. The horizontal black line indicates $S_t = 1$. Each source is plotted up to the $R_{90\%}$ radius.}
    \label{fig:st}
\end{figure*}

The fact that the smallest values of gas-to-dust mass ratio were found for Lupus discs, but the dust disc masses decrease with time, was explained in \cite{2025ApJ...989...10T} as different timescales for the evolution of the gas and the dust. The difference in gas disc mass indicates that most of its evolution happens between the ages of Ophiuchus and Lupus, with less significant changes at the later stages of evolution, traced by Lupus and UpperSco systems. Thus, the difference in gas-to-dust mass ratio between the two older regions arises entirely from the dust disc mass evolution, which is better determined in our multi-wavelength analysis, whereas the difference measured between younger systems in Ophiuchus and Lupus is additionally driven by the evolving gas component. This differential evolution highlights that the gas-to-dust ratio is not a static parameter, but a dynamic one that reflects the varying timescales of gas accretion and dust grain growth as the disc matures.

\subsection{Further constraints on the Stokes Number and Toomre Parameter}

We calculate the Stokes number, $S_t$, of the maximum grain size $a_{max}$ for our resolved sources as a function of radius, by using the properties obtained through our multi-wavelength fitting. The Stokes number is defined as: 

\begin{equation}
    S_t = \frac{\pi \rho_m a}{2\Sigma_g}
\end{equation}

where $\rho_m = 1.675$ g cm$^{-3}$ is the solid bulk density, $a$ is the grain size. In this work we use the maximum values $a_{\mathrm{max}}$, and $\Sigma_g$ is the gas surface density, estimated as $\Sigma_g = \Sigma_d/\epsilon$, where $\epsilon$ is the non-resolved dust-to-gas mass ratio computed from the densities obtained with Ricci's opacities. The results for these discs are shown in Figure \ref{fig:st}, suggest that the younger discs in Ophiuchus have grains that are well coupled with the gas ($S_t < 1$), and decoupled grains ($S_t > 1$) may only be present in the outermost part of these young discs. Similar results are obtained for Lupus, with exception of Lupus 7 and 10. The Stokes number for Lupus 7 indicate that dust grains are decoupled from the gas over the whole disc extent, while for Lupus 10, the largest grains are decoupled from the gas outside of $\sim 60$ au. Four of the UpperSco discs are also consistent with well-coupled dust grains, but UpperSco 6 and 7 show that the largest grains are decoupled from the gas in the whole extent of their disc. 

\begin{figure*}
	\includegraphics[width=\textwidth]{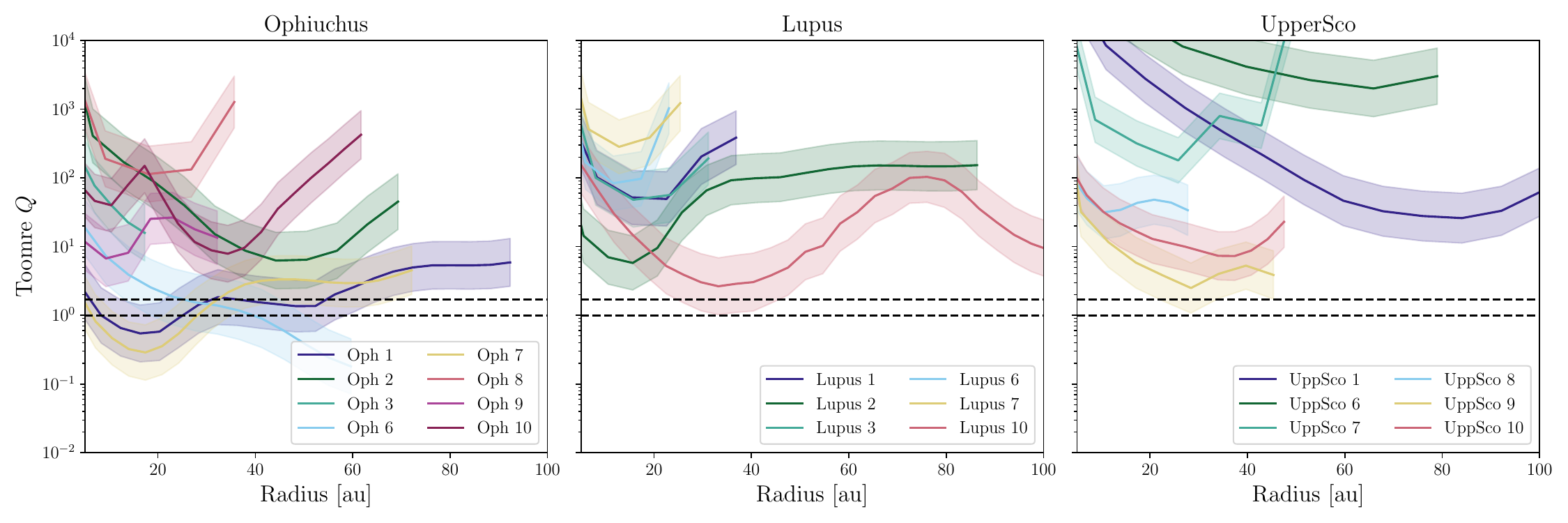}
    \caption{Toomre parameter ($Q$) dust properties obtained from our multi-wavelength analysis for the resolved discs in the AGE-PRO sample adopting Ricci's opacities. The horizontal black lines indicates $Q = 1$ and $Q=1.7$. Each source is plotted up to the $R_{90\%}$.}
    \label{fig:toomre}
\end{figure*}

Assuming the gas and dust surface density are related through the gas-to-dust ratio obtained in the previous section, we computed the Toomre $Q$ parameter, defined as

\begin{equation}
    Q = \frac{c_s \Omega_k}{\pi G \Sigma_g}
\end{equation}

where $c_s$ is the sound speed, $\Omega_k$ is the Keplerian angular velocity, and $G$ the gravitational constant. The obtained Toomre parameters when adopting Ricci's opacities are presented in Figure \ref{fig:toomre}. Note that for optically thick discs, the obtained dust (and gas) surface density represent an upper limit, and thus the Toomre parameter can be consider a lower limit for those discs.
Our results indicate that the resolved discs in the AGE-PRO sample are gravitationally stable ($Q > 1.7$) at all radii in Lupus and UpperSco regions. The Ophiuchus discs are also stable over most of their extent, but three of these young systems have regions where the disc is unstable ($Q < 1.7$): Ophiuchus 1, 7, and 6. Two are unstable in the inner $\sim 40$ au (Ophiuchus 1 and 7), while Ophiuchus 6 is unstable in the outer disc, beyond $\sim 30$ au. These three discs are massive  when compared with the rest of the sample \citep{2025ApJ...989...10T}, their respective gas and dust masses are 0.23M$_{\odot}$ and 104 M$_{\oplus}$ for Ophiuchus 1, 0.11M$_{\odot}$ and 979 M$_{\oplus}$ for Ophiuchus 6, and 0.41M$_{\odot}$ and 102 M$_{\oplus}$ for Ophiuchus 7 (computed using Ricci's opacities). For Ophiuchus 1 and 7, the dust surface density reach values of 2\,g\,cm$^{-2}$ in the innermost regions and then rapidly decreases, with a total gas-to-dust ratio of 750 and 1329 respectively, leading to low values of Toomre $Q$ in the inner disc regions. For Ophiuchus 6, the dust surface density reaches values of $\Sigma_{\rm d} > 1$\,g\,cm$^{-2}$ over the whole extent of the disc (being almost constant but with a slight increase in the outer region), which translates into the largest dust mass in the Ophiuchus region. Although its gas-to-dust ratio is not as large as for Oph 1 and 7, its large value of $\Sigma_{\rm d}$ leads to $Q<1$ in the outermost regions of Ophiuchus 6. 

An interesting result arises from the analysis done by \cite{2025ApJ...989....9V}, where the residuals from subtracting their best-fit model, which is azimuthally symmetric, identifies non-symmetric structure. For both Oph1 and Oph7, \cite{2025ApJ...989....9V} identified non-symmetric structures with high significance ($>5\sigma$). Notably, the residuals for Oph7 reveal a structure resembling a two-armed ($m=2$) spiral. A gravitationally unstable disc, such as Oph7, may develop such spiral arms, similar to those found in other systems. For example, Elias 2-27 \citep{perez2016, paneque2021} where the disc has high-contrast spiral arms and the total disc mass has been directly constrained \citep{veronesi2021} and found to be also Toomre $Q$ unstable. In the case of the younger Oph7 disc, we may be witnessing the early stages of spiral arms driven by gravitational instability (GI). The spirals detected in the Band 6 were not detected in our Band 3 and 5 observations, likely due to the lower sensitivity per beam of these observations compared to Band 6. 

It is important to remark the fact that the gas disc is expected to be larger than the dust disc, and thus simply multiplying the dust surface density by the dust-to-gas mass ratio can result in an underestimation of the Toomre parameter. We tested how different the obtained profile is
for the three discs with $Q<1$ (Ophiuchus 1, 6, and 7) in our dust analysis, if we assume that the gas surface density follows an exponentially tapered power-law function, considering the total gas mass from \cite{trapman2025} and the critical radius reported in \cite{2026MNRAS.tmp..391L}. The obtained results (Fig. \ref{fig:toomre_test}), suggest that Ophiuchus 6 is unstable in the whole extent of the disc ($Q<1$). For Ophiuchus 1 we obtained $1.7>Q>1$ for most of the disc radial extension implying a marginally stable disc. Ophiuchus 7 is consistent with the constraints made from the dust analysis with $Q<1$ for $10<r<50$, implying a gravitational unstable disc.


\section{Conclusions}\label{sec:conclusion}

In this work, we analysed multi-wavelength (3.1, 1.5, and 1.3 mm) ALMA continuum observations of the AGE-PRO survey to investigate the evolution of dust in protoplanetary discs across three regions of different ages: Ophiuchus (<1 Myr), Lupus (1–3 Myr), and Upper Scorpius (>3 Myr). By combining visibility fitting with multi-wavelength modelling, we derive spatially resolved constraints on dust disc masses and sizes, maximum grain sizes, and dynamical parameters that link the dust and gas components. The main conclusions can be summarized as follows:

\begin{itemize}
    \item The dust radii of resolved discs show no clear correlation with region age and remain consistent across all three bands (Figure \ref{fig:r90}). This lack of radial evolution suggests that radial drift is being mitigated, likely by dust traps. These results support the findings of \cite{2025ApJ...989....6K}, which suggest that the observed millimetre disc size is often defined by the position of the outermost trap.

    \item The multi-wavelength brightness radial profiles are consistent with those from \cite{2025ApJ...989....9V} and confirm that substructures are present in roughly half of the AGE-PRO sample at the current resolution. Moreover, at least three discs show signatures consistent with dust trapping in both their dust surface density and maximum grain size radial profiles. Additional observations at higher angular resolution will be necessary to confirm or discard the presence of dust traps in the remaining  discs.  

    \item The analysis per region reveals that the median spectral index ($\alpha$) is lowest in Lupus and reaches its highest values in UpperSco. Several discs exhibit $\alpha < 2$, which can be explained by optically thick emission across the observed bands. Transition discs display larger $\alpha$ compared to discs of similar sizes, which is consistent with depletion of large grains in their cavities. 

    \item The multi-wavelength fitting confirms a decline in dust mass from Ophiuchus through Lupus and UpperSco, in agreement with the single-wavelength analysis of AGE-PRO \citep{2025ApJ...989....1Z}. Although the absolute values of the derived masses depend strongly on the adopted opacity models (a factor of $\sim5-10$), the downward trend is still recovered: discs lose their dust mass over a few Myr. This is consistent with theoretical expectations of radial drift and planetesimal formation locking up solids in bodies no longer traced at millimetre wavelengths. Furthermore, in general, the dust masses obtained from the multi-wavelength modelling are larger than those obtained from single wavelength estimations.

    \item The radially averaged maximum grain size is the largest in the oldest UpperSco region, regardless of the adopted opacities. Together with enhanced dust surface densities at ring locations, these results provide evidence of pressure traps that concentrate large grains and promote their growth. 

    \item When comparing with AGE-PRO results on gas, the trend on the dust-to-gas mass ratio is consistent with those reported by \cite{2025ApJ...989...10T}, with lowest values in the Lupus region and similar values in Ophiuchus and UpperSco. This again highlights the fact that dust and gas evolve on different timescales. Stokes numbers show evidence of decoupling at large radii for a few discs, while the Toomre parameter indicates that only three massive discs in the youngest region of Ophiuchus may be gravitationally unstable: Ophiuchus 1, 6, and 7. In particular, Ophiuchus 7 (IRS 63) exhibits signatures of gravitational instability with spiral arms reported by \cite{2025ApJ...989....9V}. The detection of these features in Ophiuchus 7 reinforces the idea that GI-driven spiral arms are a transient but critical phase in the early development of young, massive protoplanetary systems.

\end{itemize}

\section*{Acknowledgements}
We thank the referee for their careful reading and valuable comments, which helped to improve the manuscript.
L.P. and C.P acknowledge support from ANID BASAL project FB210003.
L.P. acknowledges funding from ANID FONDECYT Regular \#1221442 and ANID FONDECYT Regular \#1262272.
C.P, P.P, and A.S acknowledge funding from the UK Research and Innovation (UKRI) under the UK government's Horizon Europe funding guarantee from ERC (under grant agreement No 101076489). 
CP and LL acknowledge support from the Science and Technology Facilities Council (STFC).
C.A.G. acknowledges support from FONDECYT de Postdoctorado 2021 \#3210520 and Proyectos Internos USM PI\_LII\_26\_25.
C.A.G. and L.P acknowledge support from ``Comité Mixto ESO-Gobierno de Chile 2023'', under grant 072-2023.
R.A. acknowledges funding by the European Union through the E-BEANS ERC project (grant number 100117693). 
N.K. was funded by the Deutsche Forschungsgemeinschaft (DFG, German Research Foundation) - 325594231, FOR 2634/2.

\section*{Data Availability}

This paper makes use of the following ALMA data: \\
ADS/JAO.ALMA\#2015.1.00761.S, \\
ADS/JAO.ALMA\#2019.1.01111.S, \\
ADS/JAO.ALMA\#2022.1.00638.S, \\
ALMA is a partnership of ESO (representing its member states), NSF (USA) and NINS (Japan), together with NRC (Canada), MOST and ASIAA (Taiwan), and KASI (Republic of Korea), in cooperation with the Republic of Chile. The Joint ALMA Observatory is operated by ESO, AUI/NRAO, and NAOJ. The National Radio Astronomy Observatory is a facility of the National Science Foundation operated under cooperative
agreement by Associated Universities, Inc.



\bibliographystyle{mnras}
\bibliography{example} 



\clearpage
\appendix

\section{Band 3 images}

\begin{figure*}
	\includegraphics[width=\textwidth]{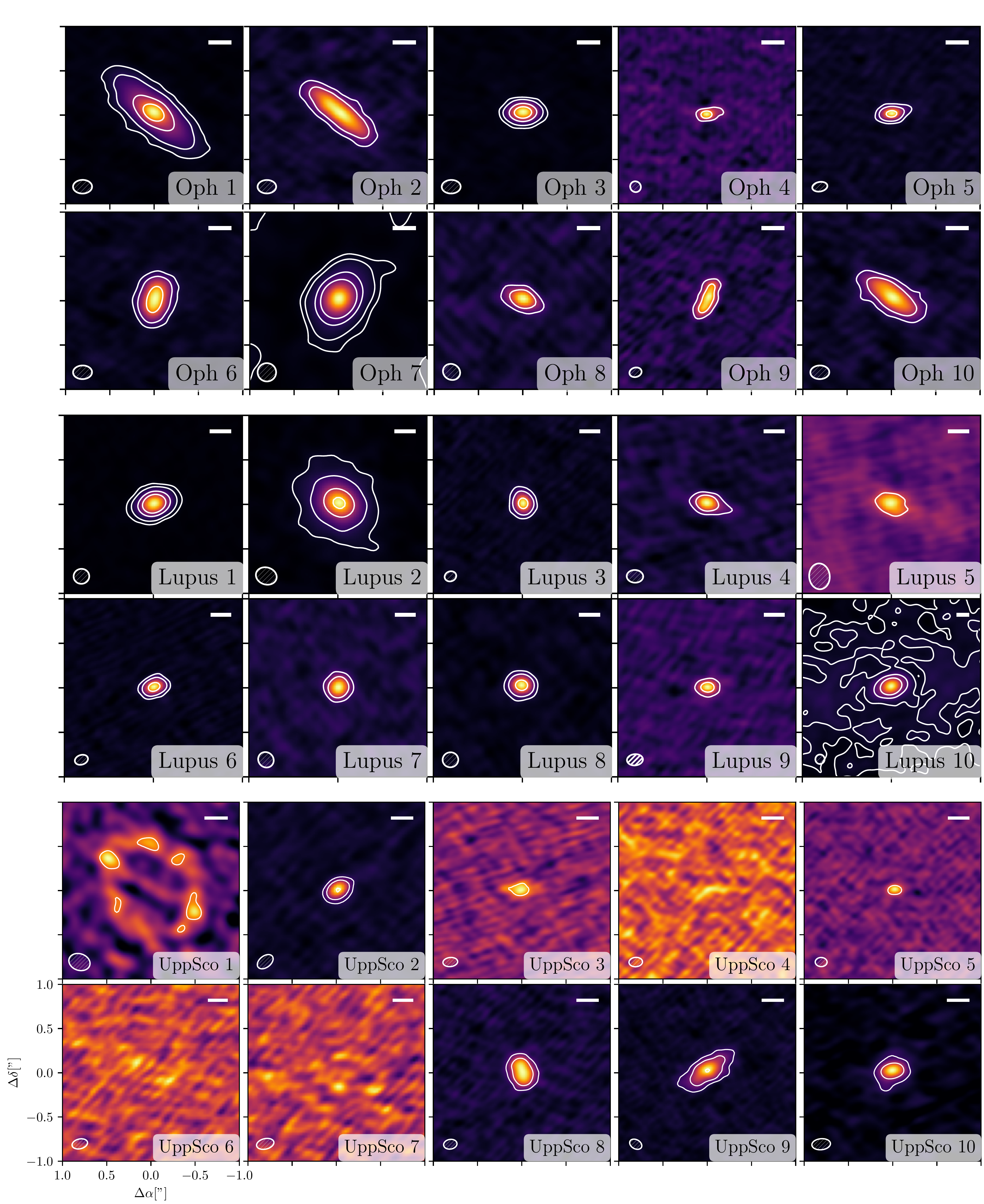}
    \caption{Band 3 (3.1 mm) continuum images of the whole AGE-PRO sample. The ellipse in the bottom left corner of each image marks the beam of the image. The total field of view (FOV) is of $2''$ for all the images except for Lupus 10, where the FOV  is of $4''$. The horizontal line in the top right of each panel correspond to a 30 au scale. The contours mark the $5\sigma$, $15\sigma$, $50\sigma$, and $95\sigma$ detections.}
    \label{fig:images_B3}
\end{figure*}

Figure \ref{fig:images_B3} shows the ALMA Band 3 continuum images of the 30 discs in the sample.

\section{Observing log}
Table \ref{tab:obs_log} shows the observing log of the Bands 3 and 5 ALMA observations used in this paper.
\begin{table*}
\centering
\caption{ALMA Observational log of our new Bands 3 (3.1 mm) and 5 (1.47 mm) observations of the whole AGE-PRO sample.}

\scriptsize
\renewcommand{\arraystretch}{1.2}
\setlength{\tabcolsep}{3pt}
\begin{tabular}{cccccccccccc}
    \hline
    \hline
    \\[0ex]
    Setup &
  Source Name &
  UTC Date &
  Config &
  \begin{tabular}[c]{@{}c@{}}Baselines\\ (m)\end{tabular} &
  N$_{ant}$ &
  \begin{tabular}[c]{@{}c@{}}Elev\\ (deg)\end{tabular} &
  \begin{tabular}[c]{@{}c@{}}Mean\\ PWV\\ (mm)\end{tabular} &
  Calibrators \\ 
  \\[0ex]
    \hline
    \\[-1ex]
    Oph B3 &
  \begin{tabular}[c]{@{}c@{}}Oph 1-3, 6-7, 10\\ Oph 1-3, 6-7, 10\\ Oph 1-3, 6-7, 10\\ Oph 4-5\\ Oph 4-5\\ Oph 8\\ Oph 9\end{tabular} &
  \begin{tabular}[c]{@{}c@{}}2021-07-27 01:01:\\ 2021-10-31 15:46:\\ 2021-11-03 19:22:\\ 2023-06-24 05:47:\\ 2023-09-26 23:21:\\ 2023-06-12 04:01:\\ 2023-06-23 03:43:\end{tabular} &
  \begin{tabular}[c]{@{}c@{}}C43-7\\ C43-7\\ C43-7\\ C43-7\\ C43-7\\ C43-7\\ C43-7\end{tabular} &
  \begin{tabular}[c]{@{}c@{}}15.0 - 3321.0\\ 63.6 - 6855.0\\ 46.8 - 5185.1\\ 84.8 - 8282.7\\ 91.6 - 8547.8\\ 79.4 - 5323.0\\ 91.6 - 8282.7\end{tabular} &
  \begin{tabular}[c]{@{}c@{}}37\\ 44\\ 43\\ 44\\ 46\\ 42\\ 45\end{tabular} &
  \begin{tabular}[c]{@{}c@{}}74.0\\ 66.5\\ 61.2\\ 41.3\\ 44.3\\ 75.0\\ 72.7\end{tabular} &
  \begin{tabular}[c]{@{}c@{}}0.6\\ 1.0\\ 1.4\\ 1.0\\ 1.2\\ 1.3\\ 1.2\end{tabular} &
  \begin{tabular}[c]{@{}c@{}}J1633-2557, J1642-2849, J1617-2537\\ J1633-2557, J1642-2849, J1617-2537\\ J1633-2557, J1642-2849, J1617-2537\\ J1617-2537, J1633-2557\\ J1617-2537, J1633-2557\\ J1617-2537, J1633-2557\\ J1617-2537, J1633-2557\end{tabular} \\ \hline
Oph B5 &
  \begin{tabular}[c]{@{}c@{}}Oph 1-3, 6-7, 10\\ Oph 4, 5, 8\\ Oph 4, 5, 8\\ Oph 4, 5, 8\\ Oph 9\end{tabular} &
  \begin{tabular}[c]{@{}c@{}}2023-05-15 04:48:\\ 2023-05-20 06:34:\\ 2023-05-21 07:38:\\ 2023-05-22 02:06:\\ 2023-04-25 06:28:\end{tabular} &
  \begin{tabular}[c]{@{}c@{}}C43-6\\ C43-6\\ C43-6\\ C43-6\\ C43-6\end{tabular} &
  \begin{tabular}[c]{@{}c@{}}15.3 - 2953.5\\ 27.1 - 3637.7\\ 27.1 - 3637.7\\ 27.1 - 3637.7\\ 15.3 - 2279.6\end{tabular} &
  \begin{tabular}[c]{@{}c@{}}45\\ 40\\ 42\\ 44\\ 42\end{tabular} &
  \begin{tabular}[c]{@{}c@{}}86.6\\ 56.7\\ 42.3\\ 62.8\\ 77.9\end{tabular} &
  \begin{tabular}[c]{@{}c@{}}23.2\\ .6\\ 0.8\\ 1.0\\ 1.3\end{tabular} &
  \begin{tabular}[c]{@{}c@{}}J1617-2537, J1633-2557\\ J1554-2704, J1553-2422, J1633-2557, J1617-2537\\ J1554-2704, J1553-2422, J1633-2557, J1617-2537\\ J1554-2704, J1553-2422, J1633-2557, J1617-2537\\ J1634-2058, J1657-2004\end{tabular} \\ \hline
Lup B3 &
 \begin{tabular}[c]{@{}c@{}}Lupus 2\\ Lupus 3\\ Lupus 4-5\\ Lupus 4-5\\ Lupus 6\\ Lupus 7\\ Lupus 8\\ Lupus 9\\ Lupus 10\end{tabular} &
  \begin{tabular}[c]{@{}c@{}}2023-06-10 03:34:\\ 2023-06-23 02:54:\\ 2023-06-11 02:40:\\ 2023-06-23 22:21:\\ 2023-06-23 00:49:\\ 2023-06-12 02:24:\\ 2023-06-12 01:35:\\ 2023-06-21 05:26:\\ 2019-06-14 02:56:\end{tabular} &
  \begin{tabular}[c]{@{}c@{}}C43-7\\ C43-7\\ C43-7\\ C43-7\\ C43-7\\ C43-7\\ C43-7\\ C43-7\\ -\end{tabular} &
  \begin{tabular}[c]{@{}c@{}}91.6 - 3697.3\\ 91.6 - 8282.7\\ 79.4 - 4614.4\\ 84.8 - 8282.7\\ 91.6 - 8282.7\\ 79.5 - 5323.0\\ 79.4 - 5323.0\\ 91.6 - 8282.7\\ 83.1 - 16196.3\end{tabular} &
  \begin{tabular}[c]{@{}c@{}}42\\ 45\\ 41\\ 44\\ 45\\ 43\\ 43\\ 43\\ 43\end{tabular} &
  \begin{tabular}[c]{@{}c@{}}75.9\\ 72.4\\ 77.1\\ 77.1\\ 66.1\\ 71.5\\ 78.3\\ 45.4\\ 71.4\end{tabular} &
  \begin{tabular}[c]{@{}c@{}}0.5\\ 1.2\\ 0.3\\ 0.7\\ 1.6\\ 1.2\\ 1.2\\ 1.6\\ 2.7\end{tabular} &
  \begin{tabular}[c]{@{}c@{}}J1523-3338, J1534-3526\\ J1610-3958, J1621-4003\\ J1523-3338, J1534-3526\\ J1523-3338, J1534-3526\\ J1610-3958, J1621-4003\\ J1604-4228, J1604-4441\\ J1534-3526, J1523-3338\\ J1621-4003, J1610-3958\\ -\end{tabular} \\ \hline
Lup B5 &
  \begin{tabular}[c]{@{}c@{}}Lupus 1-2\\ Lupus 3, 6, 8\\ Lupus 3, 6, 8\\ Lupus 4-5, 7, 9\\ Lupus 4-5, 7, 9\\ Lupus 10\end{tabular} &
  \begin{tabular}[c]{@{}c@{}}2023-05-16 05:34:\\ 2023-05-14 05:11:\\ 2023-06-08 06:05:\\ 2023-05-20 04:07:\\ 2023-05-20 05:34:\\ 2023-05-16 05:34:\end{tabular} &
  \begin{tabular}[c]{@{}c@{}}C43-6\\ C43-6\\ C43-6\\ C43-6\\ C43-6\\ C43-6\end{tabular} &
  \begin{tabular}[c]{@{}c@{}}27-1 - 2953.5\\ 15.3 - 2617.6\\ 76.6 - 3637.7\\ 27.1 - 3637.7\\ 27.1 - 3637.7\\ 27-1 - 2953.5\end{tabular} &
  \begin{tabular}[c]{@{}c@{}}42\\ 33\\ 44\\ 36\\ 41\\ 42\end{tabular} &
  \begin{tabular}[c]{@{}c@{}}70.6\\ 72.4\\ 49.2\\ 76.5\\ 66.0\\ 70.6\end{tabular} &
  \begin{tabular}[c]{@{}c@{}}2.0\\ 1.5\\ 0.7\\ 2.5\\ 2.5\\ 2.0\end{tabular} &
  \begin{tabular}[c]{@{}c@{}}J1634-3526, J1523-3338\\ J1610-3958, J1621-4003\\ J1610-3958, J1621-4003\\ J1610-3958, J1621-4003\\ J1610-3958, J1621-4003\\ J1634-3526, J1523-3338\end{tabular} \\ \hline
UppSco B3 &
  \begin{tabular}[c]{@{}c@{}}UppSco 1\\ UppSco 2\\ UppSco 3\\ UppSco 4\\ UppSco 5\\ UppSco 6\\ UppSco 7\\ UppSco 8\\ UppSco 9-10\\ UppSco 9-10\end{tabular} &
  \begin{tabular}[c]{@{}c@{}}2023-06-10 03:07:\\ 2023-06-16 01:31:\\ 2023-06-22 23:19:\\ 2023-09-28 22:36:\\ 2023-09-26 22:11:\\ 2023-06-23 05:16:\\ 2023-06-22 05:42:\\ 2023-06-23 04:29:\\ 2023-06-23 23:44:\\ 2023-09-29 20:58:\end{tabular} &
  \begin{tabular}[c]{@{}c@{}}C43-7\\ C43-7\\ C43-7\\ C43-7\\ C43-7\\ C43-7\\ C43-7\\ C43-7\\ C43-7\\ C43-7\end{tabular} &
  \begin{tabular}[c]{@{}c@{}}91.6 - 3697.3\\ 79.4 - 6855.0\\ 91.6 - 8282.7\\ 91.6 - 8547.8\\ 91.6 - 8547.8\\ 91.6 - 8282.7\\ 91.6 - 8282.7\\ 91.6 - 8282.7\\ 84.8 - 8282.7\\ 91.6 - 8547.8\end{tabular} &
  \begin{tabular}[c]{@{}c@{}}41\\ 42\\ 45\\ 48\\ 46\\ 45\\ 44\\ 45\\ 45\\ 47\end{tabular} &
  \begin{tabular}[c]{@{}c@{}}82.5\\ 78.4\\ 54.5\\ 47.9\\ 57.3\\ 47.6\\ 42.7\\ 61.2\\ 53.0\\ 69.1\end{tabular} &
  \begin{tabular}[c]{@{}c@{}}0.6\\ 1.1\\ 1.4\\ 0.8\\ 1.4\\ 0.8\\ 2.4\\ 1.0\\ 1.0\\ 1.6\end{tabular} &
  \begin{tabular}[c]{@{}c@{}}J1624-3213, J1626-2951\\ J1558-1409, J1551-1755\\ J1554-2704, J1553-2422\\ J1617-1941, J1634-2058\\ J1617-2537, J1633-2557\\ J1617-2537, J1633-2557\\ J1617-2537, J1633-2557\\ J1617-2537, J1633-2557\\ J1558-1409, J1551-1755, J1634-2058, J1617-1941\\ J1558-1409, J1551-1755, J1634-2058, J1617-1941\end{tabular} \\ \hline
UppSco B5 &
  \begin{tabular}[c]{@{}c@{}}UppSco 1\\ UppSco 2-7\\ UppSco 2-7\\ UppSco 2-7\\ UppSco 8-10\end{tabular} &
  \begin{tabular}[c]{@{}c@{}}2023-05-15 04:48:\\ 2023-05-20 06:34:\\ 2023-05-21 07:38:\\ 2023-05-22 02:06:\\ 2023-04-25 06:28:\end{tabular} &
  \begin{tabular}[c]{@{}c@{}}C43-6\\ C43-6\\ C43-6\\ C43-6\\ C43-6\end{tabular} &
  \begin{tabular}[c]{@{}c@{}}15.3 - 2953.5\\ 27.1 - 3637.7\\ 27.1 - 3637.7\\ 27.1 - 3637.7\\ 15.3 - 2279.6\end{tabular} &
  \begin{tabular}[c]{@{}c@{}}45\\ 40\\ 42\\ 44\\ 42\end{tabular} &
  \begin{tabular}[c]{@{}c@{}}86.6\\ 56.7\\ 42.3\\ 62.8\\ 77.9\end{tabular} &
  \begin{tabular}[c]{@{}c@{}}3.2\\ 2.6\\ 0.8\\ 1.0\\ 1.3\end{tabular} &
  \begin{tabular}[c]{@{}c@{}}J1617-2537, J1633-2557\\ J1554-2704, J1553-2422, J1633-2557, J1617-2537\\ J1554-2704, J1553-2422, J1633-2557, J1617-2537\\ J1554-2704, J1553-2422, J1633-2557, J1617-2537\\ J1634-2058, J1657-2004\end{tabular} \\ \hline
  \label{tab:obs_log}
\end{tabular}
\end{table*}

\FloatBarrier
\clearpage

\newpage
\section{Frankenstein's hyperparameters}

Tables \ref{tab:oph_frank_params}, \ref{tab:lup_frank_params}, and \ref{tab:upp_frank_params} show the \texttt{Frank} hyperparameters used to obtain the brightness radial profiles of the discs in our sample. The dust disc size $R_{90\%}$ is reported for each of the three bands studied (Band 6 from \cite{vioque2025}), as well as the resolution of each profile.

\begin{table*}
\centering
\caption{\texttt{Frank} hyperparameters and geometries (from \protect\cite{2025ApJ...989....9V}) used to obtain the final radial brightness profiles of the Ophiuchus disks. \texttt{Frank} FWHM represents the intrinsic resolution of the radial profile. The Band 6 measurements from \protect\cite{2025ApJ...989....9V} are included for comparison.}
\label{tab:oph_frank_params}
\begin{tabular}{@{}cccccccc@{}}
\hline
Source & 
\begin{tabular}[c]{@{}c@{}}ALMA\\Band\end{tabular} & 
\begin{tabular}[c]{@{}c@{}}PA\\{[}deg{]}\end{tabular} & 
\begin{tabular}[c]{@{}c@{}}inc\\{[}deg{]}\end{tabular} & 
$\alpha$ & 
$w_{\mathrm{smooth}}$ & 
\begin{tabular}[c]{@{}c@{}}$R_{90\%}$\\{[}au{]}\end{tabular} & 
\begin{tabular}[c]{@{}c@{}}Frank FWHM\\{[}au{]}\end{tabular} \\
\hline

Ophiuchus 1 & 
\begin{tabular}[c]{@{}c@{}}3\\5\\6\end{tabular} & 
$48.43^{+0.02}_{-0.02}$ & 
$72.59^{+0.02}_{-0.02}$ & 
\begin{tabular}[c]{@{}c@{}}1.15\\1.1\\1.05\end{tabular} & 
\begin{tabular}[c]{@{}c@{}}1e-4\\1e-4\\1e-2\end{tabular} & 
\begin{tabular}[c]{@{}c@{}}$88\pm13$\\$93\pm13$\\$95\pm14$\end{tabular} & 
\begin{tabular}[c]{@{}c@{}}12\\14\\11\end{tabular} \\
\hline

Ophiuchus 2 & 
\begin{tabular}[c]{@{}c@{}}3\\5\\6\end{tabular} & 
$50.10^{+0.07}_{-0.06}$ & 
$76.40^{+0.07}_{-0.07}$ & 
\begin{tabular}[c]{@{}c@{}}1.3\\1.3\\1.05\end{tabular} & 
\begin{tabular}[c]{@{}c@{}}1e-2\\1e-4\\1e-4\end{tabular} & 
\begin{tabular}[c]{@{}c@{}}$66\pm10$\\$66\pm10$\\$69\pm10$\end{tabular} & 
\begin{tabular}[c]{@{}c@{}}16\\15\\15\end{tabular} \\
\hline

Ophiuchus 3 & 
\begin{tabular}[c]{@{}c@{}}3\\5\\6\end{tabular} & 
$91.58^{+0.27}_{-0.28}$ & 
$61.25^{+0.25}_{-0.25}$ & 
\begin{tabular}[c]{@{}c@{}}1.30\\1.30\\1.20\end{tabular} & 
\begin{tabular}[c]{@{}c@{}}1e-1\\1e-5\\1e-1\end{tabular} & 
\begin{tabular}[c]{@{}c@{}}$17.2\pm2.4$\\$18.5\pm2.6$\\$17.5\pm2.9$\end{tabular} & 
\begin{tabular}[c]{@{}c@{}}12\\12\\11\end{tabular} \\
\hline

Ophiuchus 4 & 
\begin{tabular}[c]{@{}c@{}}3\\5\\6\end{tabular} & 
$100.7^{+2.5}_{-2.6}$ & 
$61.0^{+2.2}_{-2.2}$ & 
\begin{tabular}[c]{@{}c@{}}1.3\\1.05\\1.20\end{tabular} & 
\begin{tabular}[c]{@{}c@{}}1e-2\\1e-3\\1e-2\end{tabular} & 
\begin{tabular}[c]{@{}c@{}}$<12$\\$<10$\\$18.1\pm6.9$\end{tabular} & 
\begin{tabular}[c]{@{}c@{}}33\\32\\19\end{tabular} \\
\hline

Ophiuchus 5 & 
\begin{tabular}[c]{@{}c@{}}3\\5\\6\end{tabular} & 
$58^{+86}_{-34}$ & 
$25^{+16}_{-13}$ & 
\begin{tabular}[c]{@{}c@{}}1.2\\1.25\\1.30\end{tabular} & 
\begin{tabular}[c]{@{}c@{}}1e-4\\1e-3\\1e-1\end{tabular} & 
\begin{tabular}[c]{@{}c@{}}$<8$\\$<4.5$\\$<5.5$\end{tabular} & 
\begin{tabular}[c]{@{}c@{}}12\\15\\12\end{tabular} \\
\hline

Ophiuchus 6 & 
\begin{tabular}[c]{@{}c@{}}3\\5\\6\end{tabular} & 
$167.85^{+0.08}_{-0.08}$ & 
$74.34^{+0.12}_{-0.11}$ & 
\begin{tabular}[c]{@{}c@{}}1.1\\1.05\\1.05\end{tabular} & 
\begin{tabular}[c]{@{}c@{}}1e-2\\1e-1\\1e-4\end{tabular} & 
\begin{tabular}[c]{@{}c@{}}$40.9\pm5.9$\\$57\pm8$\\$40.2\pm5.9$\end{tabular} & 
\begin{tabular}[c]{@{}c@{}}13\\16\\12\end{tabular} \\
\hline

Ophiuchus 7 & 
\begin{tabular}[c]{@{}c@{}}3\\5\\6\end{tabular} & 
$149.57^{+0.04}_{-0.03}$ & 
$46.94^{+0.02}_{-0.03}$ & 
\begin{tabular}[c]{@{}c@{}}1.30\\1.25\\1.20\end{tabular} & 
\begin{tabular}[c]{@{}c@{}}1e-1\\1e-1\\1e-1\end{tabular} & 
\begin{tabular}[c]{@{}c@{}}$61.5\pm8.9$\\$68\pm10$\\$74\pm11$\end{tabular} & 
\begin{tabular}[c]{@{}c@{}}4.1\\12\\8.7\end{tabular} \\
\hline

Ophiuchus 8 & 
\begin{tabular}[c]{@{}c@{}}3\\5\\6\end{tabular} & 
$72.50^{+0.54}_{-0.56}$ & 
$71.84^{+0.55}_{-0.57}$ & 
\begin{tabular}[c]{@{}c@{}}1.20\\1.30\\1.05\end{tabular} & 
\begin{tabular}[c]{@{}c@{}}1e-1\\1e-1\\1e-4\end{tabular} & 
\begin{tabular}[c]{@{}c@{}}24.5\\30\\$34.5\pm7.5$\end{tabular} & 
\begin{tabular}[c]{@{}c@{}}31\\24\\20\end{tabular} \\
\hline

Ophiuchus 9 & 
\begin{tabular}[c]{@{}c@{}}3\\5\\6\end{tabular} & 
$159.55^{+0.30}_{-0.29}$ & 
$73.48^{+0.45}_{-0.43}$ & 
\begin{tabular}[c]{@{}c@{}}1.30\\1.20\\1.20\end{tabular} & 
\begin{tabular}[c]{@{}c@{}}1e-1\\1e-4\\1e-3\end{tabular} & 
\begin{tabular}[c]{@{}c@{}}$35.4\pm6.4$\\$30.6\pm4.6$\\$32.3\pm5.2$\end{tabular} & 
\begin{tabular}[c]{@{}c@{}}15\\13\\15\end{tabular} \\
\hline

Ophiuchus 10 & 
\begin{tabular}[c]{@{}c@{}}3\\5\\6\end{tabular} & 
$56.29^{+0.10}_{-0.10}$ & 
$69.53^{+0.09}_{-0.09}$ & 
\begin{tabular}[c]{@{}c@{}}1.25\\1.15\\1.20\end{tabular} & 
\begin{tabular}[c]{@{}c@{}}1e-4\\1e-2\\1e-2\end{tabular} & 
\begin{tabular}[c]{@{}c@{}}$50.9\pm7.3$\\$51.9\pm7.5$\\$62\pm10$\end{tabular} & 
\begin{tabular}[c]{@{}c@{}}18\\16\\12\end{tabular} \\
\hline

\end{tabular}
\end{table*}
\begin{table*}
\centering
\caption{\texttt{Frank} hyperparameters and geometries (from \protect\cite{2025ApJ...989....9V}) used to obtain the final radial brightness profiles of the Lupus disks. \texttt{Frank} FWHM represents the intrinsic resolution of the radial profile. The Band 6 measurements from \protect\cite{2025ApJ...989....9V} are included for comparison. In the case of Lupus 8, \texttt{Frank} failed to reproduce a radial profile.}
\label{tab:lup_frank_params}
\begin{tabular}{@{}cccccccc@{}}
\hline
Source & 
\begin{tabular}[c]{@{}c@{}}ALMA\\Band\end{tabular} & 
\begin{tabular}[c]{@{}c@{}}PA\\{[}deg{]}\end{tabular} & 
\begin{tabular}[c]{@{}c@{}}inc\\{[}deg{]}\end{tabular} & 
$\alpha$ & 
$w_{\mathrm{smooth}}$ & 
\begin{tabular}[c]{@{}c@{}}$R_{90\%}$\\{[}au{]}\end{tabular} & 
\begin{tabular}[c]{@{}c@{}}Frank FWHM\\{[}au{]}\end{tabular} \\
\hline

Lupus 1 & 
\begin{tabular}[c]{@{}c@{}}3\\5\\6\end{tabular} & 
$110.00^{+0.10}_{-0.11}$ & 
$61.41^{+0.10}_{-0.09}$ & 
\begin{tabular}[c]{@{}c@{}}1.3\\1.25\\1.2\end{tabular} & 
\begin{tabular}[c]{@{}c@{}}1e-1\\1e-1\\1e-2\end{tabular} & 
\begin{tabular}[c]{@{}c@{}}$31.3\pm0.3$\\$34.6\pm0.4$\\$34.10\pm0.95$\end{tabular} & 
\begin{tabular}[c]{@{}c@{}}27\\25\\21\end{tabular} \\
\hline

Lupus 2 & 
\begin{tabular}[c]{@{}c@{}}3\\5\\6\end{tabular} & 
$36.24^{+0.22}_{-0.22}$ & 
$38.36^{+0.13}_{-0.14}$ & 
\begin{tabular}[c]{@{}c@{}}1.15\\1.3\\1.05\end{tabular} & 
\begin{tabular}[c]{@{}c@{}}1e-3\\1e-2\\1e-4\end{tabular} & 
\begin{tabular}[c]{@{}c@{}}$85.0\pm0.5$\\$87.1\pm0.7$\\$91.1\pm1.4$\end{tabular} & 
\begin{tabular}[c]{@{}c@{}}16\\13\\18\end{tabular} \\
\hline

Lupus 3 & 
\begin{tabular}[c]{@{}c@{}}3\\5\\6\end{tabular} & 
$17.34^{+0.78}_{-0.77}$ & 
$51.83^{+0.63}_{-0.69}$ & 
\begin{tabular}[c]{@{}c@{}}1.3\\1.25\\1.05\end{tabular} & 
\begin{tabular}[c]{@{}c@{}}1e-2\\1e-1\\1e-4\end{tabular} & 
\begin{tabular}[c]{@{}c@{}}$18.9\pm2.7$\\$22.8\pm0.1$\\$31.5\pm1.2$\end{tabular} & 
\begin{tabular}[c]{@{}c@{}}7.5\\9.8\\27\end{tabular} \\
\hline

Lupus 4 & 
\begin{tabular}[c]{@{}c@{}}3\\5\\6\end{tabular} & 
$20.00\pm0.70$ & 
$31.5\pm0.60$ & 
\begin{tabular}[c]{@{}c@{}}1.2\\1.3\\1.3\end{tabular} & 
\begin{tabular}[c]{@{}c@{}}1e-3\\1e-2\\1e-1\end{tabular} & 
\begin{tabular}[c]{@{}c@{}}$<15$\\$<12$\\\textless12\end{tabular} & 
\begin{tabular}[c]{@{}c@{}}30\\15\\22\end{tabular} \\
\hline

Lupus 5 & 
\begin{tabular}[c]{@{}c@{}}3\\5\\6\end{tabular} & 
$8.2\pm2.1$ & 
$26.0\pm0.80$ & 
\begin{tabular}[c]{@{}c@{}}1.15\\1.2\\1.3\end{tabular} & 
\begin{tabular}[c]{@{}c@{}}1e-4\\1e-4\\1e-1\end{tabular} & 
\begin{tabular}[c]{@{}c@{}}$<16$\\$<20$\\\textless28\end{tabular} & 
\begin{tabular}[c]{@{}c@{}}27\\27\\28\end{tabular} \\
\hline

Lupus 6 & 
\begin{tabular}[c]{@{}c@{}}3\\5\\6\end{tabular} & 
$104.8^{+4.1}_{-4.1}$ & 
$32.2^{+2.0}_{-2.2}$ & 
\begin{tabular}[c]{@{}c@{}}1.25\\1.3\\1.05\end{tabular} & 
\begin{tabular}[c]{@{}c@{}}1e-1\\1e-4\\1e-2\end{tabular} & 
\begin{tabular}[c]{@{}c@{}}$12.8\pm1.8$\\$13.2\pm1.9$\\$21.8\pm4.1$\end{tabular} & 
\begin{tabular}[c]{@{}c@{}}9.3\\10\\17\end{tabular} \\
\hline

Lupus 7 & 
\begin{tabular}[c]{@{}c@{}}3\\5\\6\end{tabular} & 
$160.0^{+3.6}_{-3.6}$ & 
$46.7^{+2.7}_{-2.7}$ & 
\begin{tabular}[c]{@{}c@{}}1.25\\1.3\\1.05\end{tabular} & 
\begin{tabular}[c]{@{}c@{}}1e-1\\1e-1\\1e-2\end{tabular} & 
\begin{tabular}[c]{@{}c@{}}$<20$\\$<25$\\$30.4\pm3.2$\end{tabular} & 
\begin{tabular}[c]{@{}c@{}}34\\30\\22\end{tabular} \\
\hline

Lupus 8 & 
\begin{tabular}[c]{@{}c@{}}3\\5\\6\end{tabular} & 
$53.5^{+4.1}_{-4.1}$ & 
$38.9^{+2.4}_{-2.5}$ & 
\begin{tabular}[c]{@{}c@{}}-\\-\\1.3\end{tabular} & 
\begin{tabular}[c]{@{}c@{}}-\\-\\1e-1\end{tabular} & 
\begin{tabular}[c]{@{}c@{}}-\\-\\$<27$\end{tabular} & 
\begin{tabular}[c]{@{}c@{}}-\\-\\21\end{tabular} \\
\hline

Lupus 9 & 
\begin{tabular}[c]{@{}c@{}}3\\5\\6\end{tabular} & 
$159.5^{+6.9}_{-6.9}$ & 
$54.64^{+6.1}_{-6.7}$ & 
\begin{tabular}[c]{@{}c@{}}1.25\\1.15\\1.3\end{tabular} & 
\begin{tabular}[c]{@{}c@{}}1e-1\\1e-4\\1e-3\end{tabular} & 
\begin{tabular}[c]{@{}c@{}}$<16$\\$<20$\\$<31$\end{tabular} & 
\begin{tabular}[c]{@{}c@{}}22\\28\\39\end{tabular} \\
\hline

Lupus 10 & 
\begin{tabular}[c]{@{}c@{}}3\\5\\6\end{tabular} & 
$107.87^{+0.06}_{-0.06}$ & 
$52.02^{+0.04}_{-0.04}$ & 
\begin{tabular}[c]{@{}c@{}}1.2\\1.25\\1.05\end{tabular} & 
\begin{tabular}[c]{@{}c@{}}1e-2\\1e-1\\1e-4\end{tabular} & 
\begin{tabular}[c]{@{}c@{}}$291\pm4$\\$302\pm4$\\$353\pm2.2$\end{tabular} & 
\begin{tabular}[c]{@{}c@{}}10\\14\\14\end{tabular} \\
\hline

\end{tabular}
\end{table*}

\begin{table*}
\centering
\caption{\texttt{Frank} hyperparameters and geometries (from \protect\cite{2025ApJ...989....9V}) used to obtain the final radial brightness profiles of the Upper Scorpius disks. \texttt{Frank} FWHM represents the intrinsic resolution of the radial profile. The Band 6 measurements from \protect\cite{2025ApJ...989....9V} are included for comparison. In the case of UpperSco 3, 4, and 5, \texttt{Frank} failed to reproduce a radial profile due to the low SNR.}
\label{tab:upp_frank_params}
\begin{tabular}{@{}cccccccc@{}}
\hline
Source & 
\begin{tabular}[c]{@{}c@{}}ALMA\\Band\end{tabular} & 
\begin{tabular}[c]{@{}c@{}}PA\\{[}deg{]}\end{tabular} & 
\begin{tabular}[c]{@{}c@{}}inc\\{[}deg{]}\end{tabular} & 
$\alpha$ & 
$w_{\mathrm{smooth}}$ & 
\begin{tabular}[c]{@{}c@{}}$R_{90\%}$\\{[}au{]}\end{tabular} & 
\begin{tabular}[c]{@{}c@{}}Frank FWHM\\{[}au{]}\end{tabular} \\
\hline

UppSco 1 & 
\begin{tabular}[c]{@{}c@{}}3\\5\\6\end{tabular} & 
$45.10^{+0.20}_{-0.90}$ & 
$37.00^{+0.10}_{-0.20}$ & 
\begin{tabular}[c]{@{}c@{}}1.1\\1.3\\1.05\end{tabular} & 
\begin{tabular}[c]{@{}c@{}}1e-2\\1e-2\\1e-4\end{tabular} & 
\begin{tabular}[c]{@{}c@{}}$83.4\pm2.3$\\$94.1\pm1.1$\\$87.64\pm0.62$\end{tabular} & 
\begin{tabular}[c]{@{}c@{}}30\\24\\14\end{tabular} \\
\hline

UppSco 2 & 
\begin{tabular}[c]{@{}c@{}}3\\5\\6\end{tabular} & 
$49.4^{+2.7}_{-2.5}$ & 
$56.4^{+2.6}_{-2.3}$ & 
\begin{tabular}[c]{@{}c@{}}1.2\\1.1\\1.3\end{tabular} & 
\begin{tabular}[c]{@{}c@{}}1e-4\\1e-1\\1e-1\end{tabular} & 
\begin{tabular}[c]{@{}c@{}}$20.8\pm0.1$\\$19.1\pm0.1$\\$<17$\end{tabular} & 
\begin{tabular}[c]{@{}c@{}}19\\17\\21\end{tabular} \\
\hline

UppSco 3 & 
\begin{tabular}[c]{@{}c@{}}3\\5\\6\end{tabular} & 
$78.1\pm9.6$ & 
$58.20\pm0.13$ & 
\begin{tabular}[c]{@{}c@{}}-\\-\\1.2\end{tabular} & 
\begin{tabular}[c]{@{}c@{}}-\\-\\1e-3\end{tabular} & 
\begin{tabular}[c]{@{}c@{}}-\\-\\$27.0\pm6.6$\end{tabular} & 
\begin{tabular}[c]{@{}c@{}}-\\-\\32\end{tabular} \\
\hline

UppSco 4 & 
\begin{tabular}[c]{@{}c@{}}3\\5\\6\end{tabular} & 
$69^{+18}_{-11}$ & 
$79.3^{+7.5}_{-18.8}$ & 
\begin{tabular}[c]{@{}c@{}}-\\-\\1.05\end{tabular} & 
\begin{tabular}[c]{@{}c@{}}-\\-\\1e-4\end{tabular} & 
\begin{tabular}[c]{@{}c@{}}-\\-\\$<56$\end{tabular} & 
\begin{tabular}[c]{@{}c@{}}-\\-\\53\end{tabular} \\
\hline

UppSco 5 & 
\begin{tabular}[c]{@{}c@{}}3\\5\\6\end{tabular} & 
$139.0\pm5.3$ & 
$15\pm10$ & 
\begin{tabular}[c]{@{}c@{}}-\\-\\1.3\end{tabular} & 
\begin{tabular}[c]{@{}c@{}}-\\-\\1e-1\end{tabular} & 
\begin{tabular}[c]{@{}c@{}}-\\-\\$<19$\end{tabular} & 
\begin{tabular}[c]{@{}c@{}}-\\-\\20\end{tabular} \\
\hline

UppSco 6 & 
\begin{tabular}[c]{@{}c@{}}3\\5\\6\end{tabular} & 
$59.8^{+2.5}_{-2.5}$ & 
$62.9^{+2.6}_{-2.7}$ & 
\begin{tabular}[c]{@{}c@{}}1.1\\1.1\\1.05\end{tabular} & 
\begin{tabular}[c]{@{}c@{}}1e-2\\1e-3\\1e-4\end{tabular} & 
\begin{tabular}[c]{@{}c@{}}$56.2\pm4.2$\\$79\pm3$\\$63\pm19$\end{tabular} & 
\begin{tabular}[c]{@{}c@{}}40\\32\\37\end{tabular} \\
\hline

UppSco 7 & 
\begin{tabular}[c]{@{}c@{}}3\\5\\6\end{tabular} & 
$179^{+11}_{-11}$ & 
$32.4^{+6.2}_{-9.0}$ & 
\begin{tabular}[c]{@{}c@{}}1.15\\1.3\\1.2\end{tabular} & 
\begin{tabular}[c]{@{}c@{}}1e-2\\1e-4\\1e-2\end{tabular} & 
\begin{tabular}[c]{@{}c@{}}$50.9\pm0.4$\\$46.8\pm0.5$\\$47\pm22$\end{tabular} & 
\begin{tabular}[c]{@{}c@{}}27\\26\\21\end{tabular} \\
\hline

UppSco 8 & 
\begin{tabular}[c]{@{}c@{}}3\\5\\6\end{tabular} & 
$16.27^{+0.32}_{-0.33}$ & 
$56.24^{+0.28}_{-0.29}$ & 
\begin{tabular}[c]{@{}c@{}}1.3\\1.05\\1.05\end{tabular} & 
\begin{tabular}[c]{@{}c@{}}1e-1\\1e-1\\1e-4\end{tabular} & 
\begin{tabular}[c]{@{}c@{}}$26.8\pm0.3$\\$28.8\pm0.3$\\$27.19\pm0.18$\end{tabular} & 
\begin{tabular}[c]{@{}c@{}}12\\15\\15\end{tabular} \\
\hline

UppSco 9 & 
\begin{tabular}[c]{@{}c@{}}3\\5\\6\end{tabular} & 
$125.40^{+0.10}_{-0.09}$ & 
$75.10^{+0.12}_{-0.13}$ & 
\begin{tabular}[c]{@{}c@{}}1.1\\1.3\\1.05\end{tabular} & 
\begin{tabular}[c]{@{}c@{}}1e-1\\1e-1\\1e-4\end{tabular} & 
\begin{tabular}[c]{@{}c@{}}$47.9\pm0.6$\\$46.7\pm0.3$\\$47.80\pm0.34$\end{tabular} & 
\begin{tabular}[c]{@{}c@{}}8.8\\16\\20\end{tabular} \\
\hline

UppSco 10 & 
\begin{tabular}[c]{@{}c@{}}3\\5\\6\end{tabular} & 
$155.73^{+0.26}_{-0.26}$ & 
$49.14^{+0.22}_{-0.21}$ & 
\begin{tabular}[c]{@{}c@{}}1.2\\1.3\\1.05\end{tabular} & 
\begin{tabular}[c]{@{}c@{}}1e-1\\1e-2\\1e-3\end{tabular} & 
\begin{tabular}[c]{@{}c@{}}$41.3\pm0.7$\\$30.5\pm0.2$\\$42.0\pm1.5$\end{tabular} & 
\begin{tabular}[c]{@{}c@{}}12\\11\\12\end{tabular} \\
\hline

\end{tabular}
\end{table*}

\FloatBarrier
\clearpage

\newpage
\section{Multi-wavelength fitting: Obtained profiles}

\begin{figure*}
	\includegraphics[width=\textwidth]{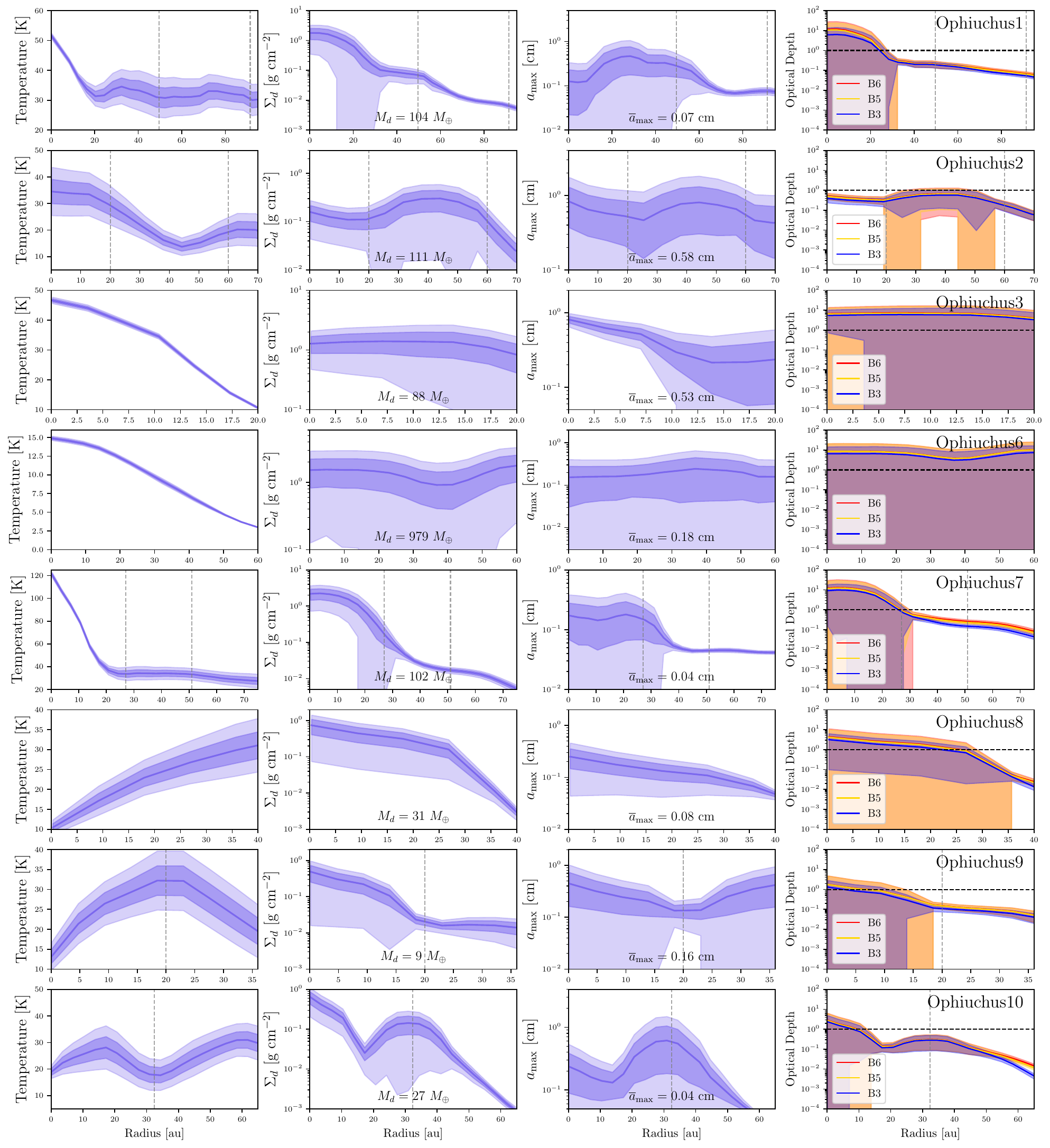}
    \caption{Dust properties profiles obtained with the SED fitting for the resolved sources in Ophiuchus using Ricci's opacities. The error bars are computed from the percentiles 32 and 68. All discs are plotted up to $R_{90\%}$, and the vertical lines represent the locations of the detected ring-like substructures.}
    \label{fig:ophiuchus}
\end{figure*}

\begin{figure*}
	\includegraphics[width=\textwidth]{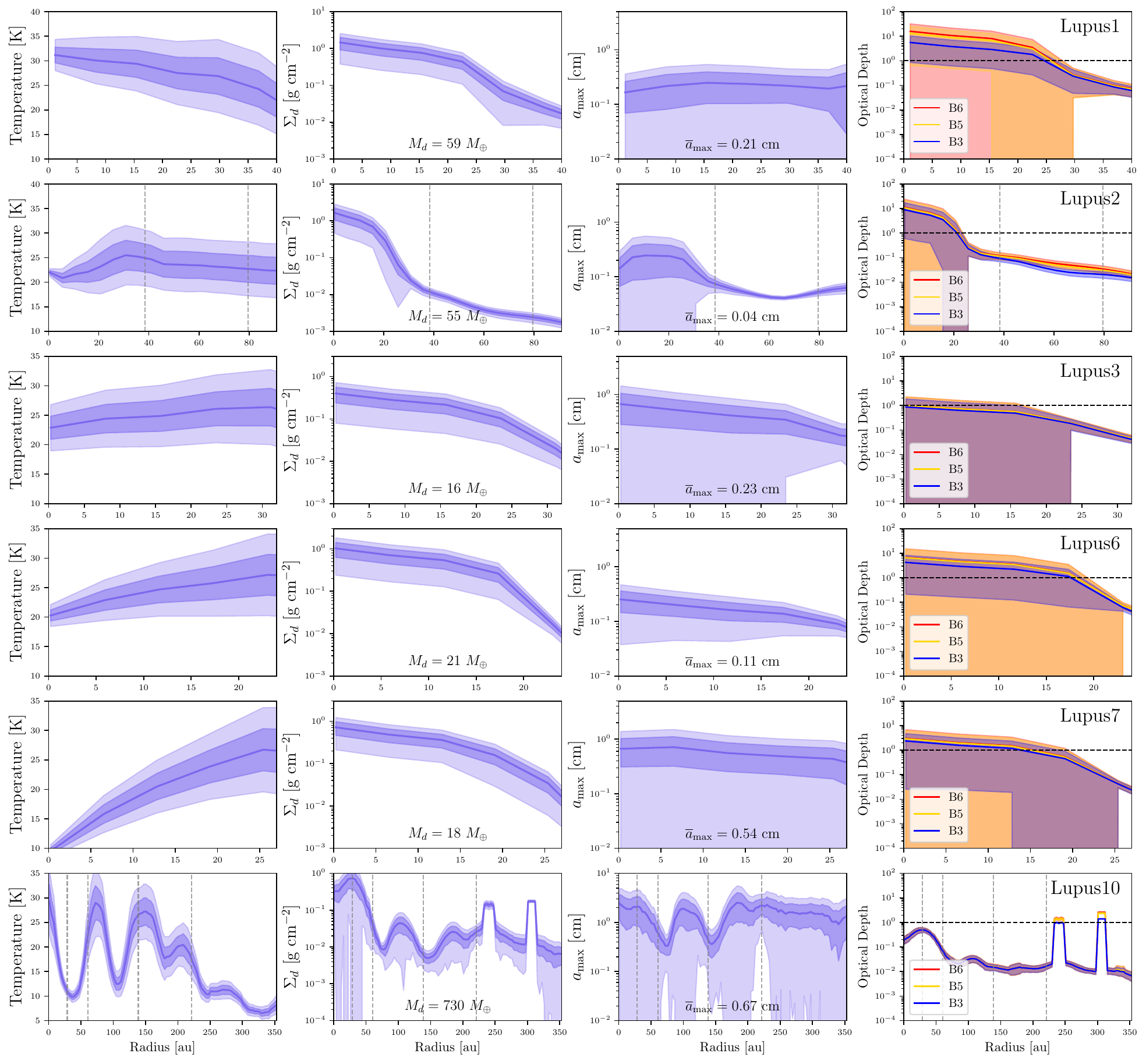}
    \caption{Same as \ref{fig:ophiuchus} but for Lupus discs.}
    \label{fig:lupus}
\end{figure*}

\begin{figure*}
	\includegraphics[width=\textwidth]{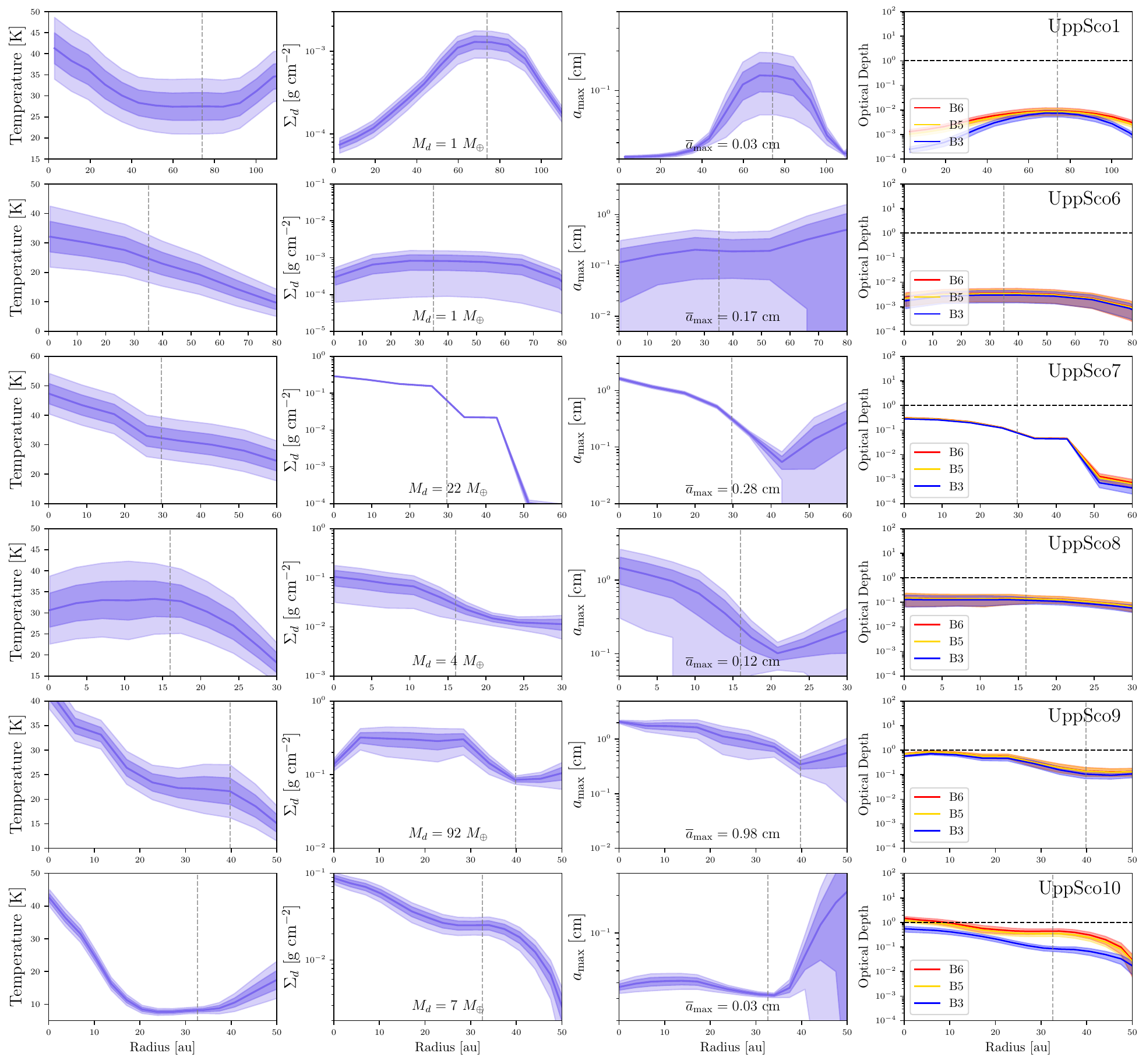}
    \caption{Same as \ref{fig:ophiuchus} but for UpperSco discs.}
    \label{fig:uppsco}
\end{figure*}

The Figures in this section (\ref{fig:ophiuchus}, \ref{fig:lupus}, and \ref{fig:uppsco}) show the obtained radial profiles for the dust disc temperature, dust surface density, and maximum grain size for the 20 resolved discs in our sample using Ricci's opacities. The last column shows also the estimated optical depth for the three studied ALMA bands.

\section{Constraints on maximum grain size}

\begin{figure*}
	\includegraphics[width=\textwidth]{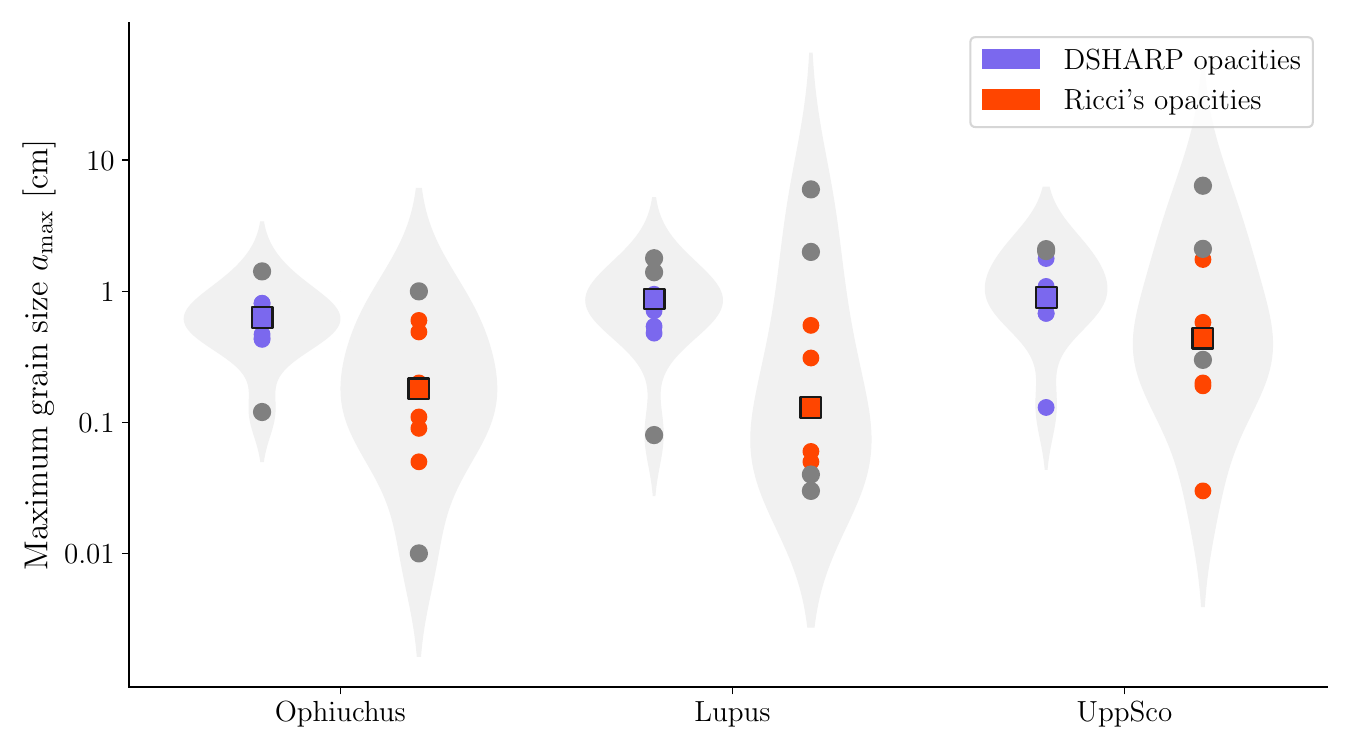}
    \caption{Average maximum grain size, $\langle a_{\mathrm{max}}\rangle$, per region obtained from the SED fitting using DSHARP (in purple), and Ricci's (in purple) opacities. Grey points marks the results for unresolved sources.}
    \label{fig:amax}
\end{figure*}

A comparison of the obtained maximum grain size for each disc of the sample using Ricci's and DSHARP opacities is shown in Figure \ref{fig:amax}.

\section{Toomre parameter}

\begin{figure*}
	\includegraphics[width=0.6\textwidth]{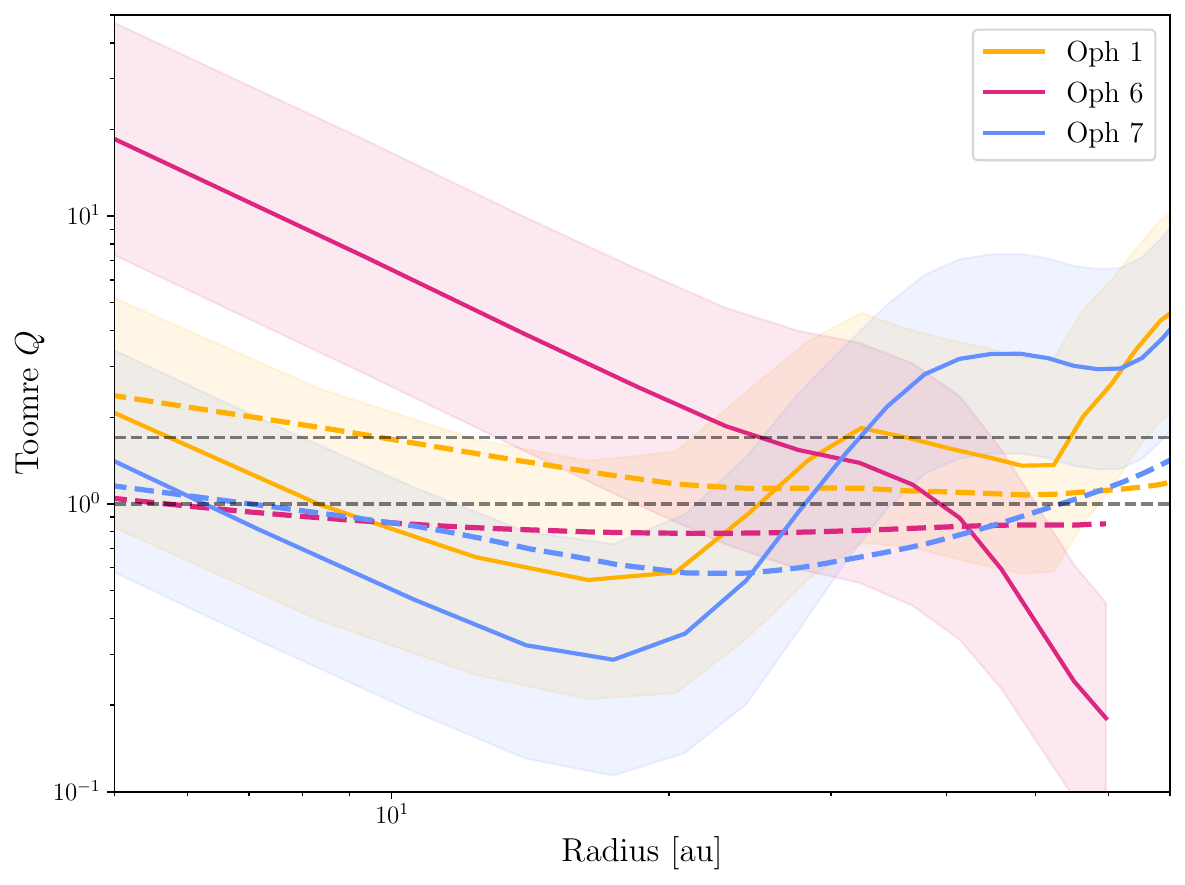}
    \caption{Comparison of the Toomre parameter $Q$ obtained multiplying the dust surface density from our analysis by the dust-to-gas mass ratio (solid lines), and the one obtained by assuming an exponentially tapered power-law profile for the gas surface density, estimated from the total gas mass and radius reported in \protect\cite{trapman2025} and \protect\cite{2026MNRAS.tmp..391L}, respectively (dashed lines) for the three discs with $Q<1$.}
    \label{fig:toomre_test}
\end{figure*}

Figure \ref{fig:toomre_test} shows the differences in the obtained Toomre parameter profile when assuming an exponentially tapered power-law profile for the gas surface density, and when simply multiplying the dust surface density profile by the dust-to-gas mass ratio for the three discs spotted as potentially unstable from our dust analysis (Oph1, Oph6, and Oph7).

\bsp	
\label{lastpage}
\end{document}